# Tangent space functional reconfigurations in individuals at risk for alcohol use disorder

Mahdi Moghaddam[1,2], Mario Dzemidzic[3,4], Daniel Guerrero[1,2], Mintao Liu[1,2], Jonathan Alessi[3,4], Martin H. Plawecki[4,5], Jaroslaw Harezlak[4,6], David A. Kareken[3,4,5]*, Joaquín Goñi[1,2,4,7]*

[1]School of Industrial Engineering, Purdue University, West-Lafayette, IN, USA.

[2]Purdue Institute for Integrative Neuroscience, Purdue University, West-Lafayette, IN, USA.

[3]Department of Neurology, Indiana University School of Medicine, Indianapolis, IN, USA

[4]Indiana Alcohol Research Center, Indiana University School of Medicine, Indianapolis, IN, USA

[5]Department of Psychiatry, Indiana University School of Medicine, Indianapolis, IN USA

[6]Department of Epidemiology and Biostatistics, Indiana University Bloomington, Bloomington, IN, USA.

[7]Weldon School of Biomedical Engineering, Purdue University, West-Lafayette, IN, USA.

*both authors equally contributed.

## Abstract

Human brain function dynamically adjusts to ever-changing stimuli from the external environment. Studies characterizing brain functional reconfiguration are nevertheless scarce. Here we present a principled mathematical framework to quantify brain functional reconfiguration when engaging and disengaging from a stop signal task (SST). We apply tangent space projection (a Riemannian geometry mapping technique) to transform functional connectomes (FCs) of 54 participants and quantify functional reconfiguration using the correlation distance of the resulting tangent-FCs. Our goal was to compare functional reconfigurations in individuals at risk for alcohol use disorder (AUD). We hypothesized that functional reconfigurations when transitioning to/from a task would be influenced by family history of alcohol use disorder (FHA) and other AUD risk factors. Multilinear regression models showed that engaging and disengaging functional reconfiguration were associated with FHA and recent drinking. When *engaging* in the SST after a rest condition, functional reconfiguration was negatively associated with recent drinking, while functional reconfiguration when *disengaging* from the SST was negatively associated with FHA. In both models, several other factors contributed to the functional reconfiguration. This study demonstrates that tangent-FCs can characterize task-induced functional reconfiguration, and that it is related to AUD risk.

## Author Summary

Human brain function constantly adapts to the external environment stimuli and transitions between cognitive states, which can be hindered by alcohol misuse and family history of alcohol use disorder (AUD). In this work, we used a novel methodology that applies Riemannian Geometry concepts to functional connectivity to quantify functional reconfiguration of rest-to-task (engaging) and task-to-rest (disengaging) transitions. We ultimately aimed to determine the relationship between AUD risk factors and functional reconfiguration. Our findings showed that engaging functional reconfiguration was diminished in participants with family history of AUD, whereas disengaging functional reconfiguration was diminished with greater recent drinking behavior. This study suggests that analysis of functional reconfiguration using Riemannian Geometry is a promising avenue to better understand rest-to-task and task-to-rest brain transitions.

## 1 Introduction

Brain functional connectivity has been used to predict behavioral attributes with the goal of understanding the relationship between individual characteristics (cognition, behavior, etc.) and brain functional networks (Amico et al., 2020; Eichenbaum et al., 2021; Fornito et al., 2015; Pervaiz et al., 2020; Van Der Wijk et al., 2022). These studies typically use data from fMRI scans that are acquired under one specific condition (e.g., during a quiet rest or while executing a motor or demanding attentional task).

Resting-state (task-free) fMRI design is now a default condition in which to assess functional connectivity (Smitha et al., 2017), such as while participants fixate on a crosshair or closing their eyes, each without other specific cognitive demands. Functional connectomes (FCs) during rest have been used as predictors of various traits including intelligence (Finn et al., 2015; Hearne et al., 2016), attention (Rosenberg et al., 2015), impulsivity (Li et al., 2013), and cognitive deficits (Svaldi et al., 2021). They have also shown test-retest reliability in the form of functional connectivity *fingerprints* (Abbas et al., 2021, 2023; Amico & Goñi, 2018; Finn et al., 2015; Venkatesh et al., 2020)- i.e., as reflective of individual identity. The resting state design is also well-suited for multi-site (Bari et al., 2019) and longitudinal studies as it is independent of participants' performance capabilities, attention span, age, or other specific limitations. (Finn et al., 2017; Rosenberg et al., 2015).

On the other hand, the resting state can also be considered as an unconstrained brain state, influenced by level of wakefulness, affective state (e.g., mood, anxiousness, etc.) and other factors (Betzel et al., 2017; Buckner et al., 2013; Finn, 2021), including activities immediately preceding the resting state scan (Chen et al., 2018; Tung et al., 2013). Others have therefore argued that functional connectivity from active cognitive engagement is a more controlled and specific probe that can better predict traits and behavior (Finn & Bandettini, 2021; Greene et al., 2018; Jiang et al., 2020; Zhao et al., 2023) and be more easily interpreted (Finn, 2021).

Complementing both views, our group is interested in the "between states" functional reconfiguration involved in both the rise to task engagement as well as the task's lingering after-effects (Amico et al., 2020; Barnes et al., 2009; Chen et al., 2018). That is, transitions or functional reconfiguration between rest and task occur gradually and may vary across participants and clinical populations. Specifically, we are interested in brain endophenotypic risk from a family history of alcohol use disorder (AUD) and how it affects transitions to and from task demands. Those with a family history of alcohol use disorder (FHA) have an approximate doubled risk for future AUD (Nurnberger et al., 2004). Insofar as FHA also affects executive cognitive ability (Braun et al., 2015; Gallen et al., 2016; Nurnberger et al., 2004; Schultz & Cole, 2016; Shine et al., 2019), it would seem likely that brain network flexibility would be affected.

Most FHA studies examine functional connectivity between a pre-defined region (a "seed") and other areas, either at rest or during cognitive tasks (Cservenka, Casimo, et al., 2014; Cservenka, Fair, et al., 2014; Herting et al., 2011; Martz et al., 2019; Weiland et al., 2013; Wetherill et al., 2012). The findings are not entirely consistent but suggest greater ventral striatal-to-frontal functional connectivity in FHA (although the frontal regions vary) and less neo-cortical functional connectivity. Whole brain analyses of FHA with quantitative metrics that assess large-scale network organization across the whole brain are scarce. Two such network-level studies of adolescents found that frontal and premotor areas showed FHA-related changes in time-averaged resting-state functional connectivity that relate to impulsivity/externalizing symptoms and psychomotor speed (Holla et al., 2017; Vaidya et al., 2019). More recently, Elton et al., (2021) identified risk-related resting state functional connectivity between pair-wise nodes in healthy subjects whose siblings had either DSM-IV alcohol abuse or dependence, finding canonical correlations between (predominantly) fronto-parietal nodes and cognitive and behavioral factors. Unusually, however, greater intelligence was related to the risk-associated connectivity differences in FHA.

In contrast, we previously studied functional connectivity immediately following a task in those with and without FHA, using connICA, an independent component analysis framework (Amico et al., 2017), to dissect FCs into individual components of connectivity (Amico et al., 2020). It was found that individuals without FHA featured most prominently in a particular functional connectivity component during the post-task rest period.

FCs are frequently represented as matrices computed using Pearson correlation coefficients between the time series of brain region pairs. FCs lie on a symmetric positive definite (SPD) manifold and are bound by its geometry (Dadi et al., 2019; You & Park, 2021). Therefore, their elements are inherently inter-related (You & Park, 2021) which may reduce the accuracy of predictions and associations (Tian & Zalesky, 2021) by violating "the uncorrelated feature assumption" (Ng et al., 2016). To overcome this problem, recent studies have instead proposed the use of tangent space projections of FCs (i.e. tangent-FCs) (Abbas et al., 2023; Pervaiz et al., 2020; You & Park, 2021), an application of Riemannian geometry (Pennec et al., 2006). The purpose of tangent space projection of FCs is to map the FCs to a Euclidean space, which removes the inter-relatedness of functional couplings. This is accomplished by first computing a reference matrix, $C_{ref}$, on the SPD manifold (Pervaiz et al., 2020). For every $C_{ref}$, the derivatives of the curves crossing $C_{ref}$ on the manifold form a Euclidean space which is tangent to $C_{ref}$ (You & Park, 2021), i.e., the tangent space of $C_{ref}$. In this study, we computed $C_{ref}$ as the Riemann mean of the FCs. Subsequently, we can compute tangent-FCs using the derivative of the curves that connect FCs to $C_{ref}$ on the manifold. Tangent-FCs have been shown to be more reliable in finding associations with individual or demographic traits and conditions as compared to FCs (Dadi et al., 2019; Dodero et al., 2015; Ng et al., 2017; Qiu et al., 2015; Wong et al., 2018). More recently, Abbas et al., (2023) showed that tangent-FCs also carry a more precise fingerprint and Simeon et al., (2022) showed their application in to harmonizing multi-site data.

The aim of this work was to utilize tangent-FCs to determine the relationship between AUD risk factors and functional reconfiguration as individuals engage in and disengage from a task. Briefly, computation of tangent-FCs requires (i) estimation of FCs, (ii) regularization of FCs, and (iii) computation of a reference matrix for tangent space projection. Our data were acquired in two consecutive fMRI scans. In the first scan the participants underwent 8 minutes of rest, while in the second scan (immediately after) participants performed 4 minutes of SST followed by 8 minutes of rest. We estimated functional connectivity after dividing the two scans into five 4-minute segments (Figure 1). We then regularized the resulting FCs and used the Riemann mean of the first resting state segment as the reference matrix for tangent space projection (see Section 2.4 for definitions). We then show that the choice of different resting state fMRI segments to compute the reference minimally affects functional reconfiguration estimates (Section 3.1). We also tested a range of regularization values and assessed their impact on the range and variance of tangent-FC elements which in turn can influence their predictive and explanation power (Section 3.6)

Finally, we used the correlation distance of tangent-FCs to measure functional reconfiguration from rest-to-task (engaging) and task-to-rest (disengaging). We hypothesized that task-to-rest and rest-to-task functional reconfiguration would be associated with AUD risk factors. We used multilinear regression analysis with engaging and disengaging functional reconfiguration as response variables to evaluate our hypothesis.

# 2 Materials and Methods

## 2.1 Participant information

The fMRI study data used here were previously reported by Amico et al., (2020) and consist of 54 participants (Table 1), among whom twenty-three were FHA positive, defined as having at least one first degree relative with a history of AUD. FHA negative participants had no first- or second-degree relatives with a history of AUD. We determined family history through interviews using the family history module of the Semi-Structured Assessment for the Genetics of Alcoholism (Bucholz et al., 1994). All participants signed an informed consent prior to study procedures, all of which were approved by the Indiana University Institutional Review Board.

Three FHA positive participants had affected mothers but reported that their mothers did not drink during pregnancy. These three participants had two to four years of college education, had no obvious facial abnormalities as reported by the interviewing technicians, and had go and stop signal reaction times within one standard deviation of the remainder of the sample.

**Table 1**

Participant characteristics

| | FHA positive (n=23, 9 males) | | FHA negative (n=31, 16 males) | | Full cohort (n=54, 25 males) | |
|---|---|---|---|---|---|---|
| | Mean (SD) | Range | Mean (SD) | Range | Mean (SD) | Range |
| Age | 23.04 (1.64) | 21 – 26 | 22.35 (1.58) | 21 – 26 | 22.64 (1.62) | 21 - 26 |
| Education | 15.32 (1.22) | 13 – 18 | 15.23 (1.20) | 14 – 19 | 15.25 (1.20) | 13 - 19 |
| SSRT[a] (ms) | 250 (48) | 157 – 397 | 230 (52) | 124 – 346 | 238 (51) | 124 - 397 |
| CES-D[b] | 8.04 (5.52)[d] | 1 – 24 | 4.41 (4.21)[d] | 0 – 17 | 5.96 (5.09) | 0 - 24 |
| AUDIT[c] | 10.26 (6.52) | 2 – 29 | 7.26 (3.92) | 1 – 20 | 8.53 (5.35) | 1 – 29 |
| Drinking days | 16.17 (8.02) | 6 – 33 | 10.54 (5.60) | 3 – 25 | 12.94 (7.24) | 3 - 33 |
| Drinks per week | 12.27 (11.45) | 2 – 51 | 7.11 (4.91) | 1 – 21 | 9.31 (8.64) | 1.20 – 51.40 |
| Drinks per drinking day | 3.66 (1.92) | 1 – 9 | 3.67 (2.39) | 1 - 10 | 3.67 (2.18) | 1.30 – 10.40 |

[a] Stop Signal Reaction Time; time to withdraw a response.

[b] Center for Epidemiologic Studies Depression scale. Scores of 16 or greater indicate a risk for clinical depression (Radloff, 1977).

[c] Alcohol Use Disorder Identification Test (Saunders et al., 1993). A score of 8 or above suggests hazardous or harmful alcohol use.

[d] Significant difference between FHA positive and negative participants ($p = 0.008$).

## 2.2 Stop Signal Task

As described in (Amico et al., 2020), the Stop Signal Task (SST) consisted of 54 Go trials and 26 Stop trials, with brief practice (8 Go trials, 7 stop trials) before imaging. Participants were instructed to respond as quickly and accurately as possible by pressing a right or left button on an MRI-compatible button box (Current Designs, Philadelphia, PA) that correspond to the Go signal of right or left pointing horizontal blue arrows. Stop trials (signaled by a red up-pointing arrow immediately following a Go stimulus) required participants to inhibit their Go response. An adaptive staircase algorithm adjusted the delay between Go and Stop stimuli in 50 ms increments to help assure the 50% accuracy rate needed to calculate stop signal reaction time (SSRT). This task was programmed using E-Prime 2.0 software (Psychology Software Tools Inc., Sharpsburg, PA). A participant's estimated SSRT was calculated according to Band et al. (2003), by subtracting the average stop-signal delay from that participant's xth percentile Go response time, where x corresponds to the stop failure rate. Participants viewed stimuli back-projected on a screen at the rear of the scanner bore. Briefly, a lower SSRT reflects faster inhibition a previously initiated response (Eagle et al., 2008). Participants' performance during SST is presented in Table 2.

**Table 2**

Stop signal task performance

| | | SSRT (ms) | %Cor. Go | GoRT (ms) | DelayDur (ms) | Stop Failure (%) |
|---|---|---|---|---|---|---|
| Full cohort | Average | 238 | 0.95 | 454 | 190 | 49% |
| (n=54) | SD | 51 | 0.08 | 91 | 99 | 7% |
| FHA positive | Average | 250 | 0.94 | 453 | 177 | 50% |
| (n=23) | SD | 48 | 0.08 | 88 | 94 | 7% |
| FHA negative | Average | 230 | 0.95 | 455 | 199 | 48% |
| (n=31) | SD | 52 | 0.09 | 95 | 103 | 7% |

%Cor. Go= Percent Go Correct; GoRT= Go Reaction Time, DelayDur= average time between Go and Stop stimulus, and Stop Failure= percentage of stop trials on which subjects made a Go response.

## 2.3 MRI acquisition and preprocessing

Participants were imaged in a 3T Siemens Prisma MRI scanner with a 64-channel head coil (neck elements off). Functional MRI data were acquired with a blood oxygenation level dependent (BOLD) contrast sensitive sequence (gradient-echo echo-planar imaging (GE-EPI), multiband (MB) slice acceleration factor 3, repetition/echo time TR/TE= 1,200 ms/29 ms, flip angle 65°, 2.5×2.5×2.5 $mm^3$ voxels, 220×220 $mm^2$ field of view, 54 interleaved axial slices). During the first BOLD fMRI scan (8:00 min; 400 volumes) participants were at rest and instructed to fixate their gaze on a central white crosshair. The next fMRI scan (12:12 min, 610 volumes) included 4 min of SST performance, followed by a short 12 s transition period when a slide announced an upcoming 8 min rest (Fig 1A). Initial 7 s in each scan (calibrations; MR signal reaching steady state magnetization) were excluded from subsequent analyses.

Participants first received a whole-brain T1-weighted structural MRI (3D Magnetization Prepared Rapid Gradient Echo (MPRAGE) sequence; duration 5:12 min, 176 sagittal slices, 1.1×1.1×1.2 $mm^3$ voxels, GRAPPA R=2 acceleration). In addition, two short (16 s) spin echo EPI scans (TR/TE= 1560/49.8 ms, five in A-P and five in P-A direction, same imaging volume/voxel size as for the GE-EPI scans) were acquired immediately before each BOLD fMRI scan. These phase-reversed spin echo EPI scans provided field maps to correct EPI geometric distortion using FSL's topup/applytopup (Smith et al., 2004).

The analysis pipeline originally applied to this dataset in Amico et al., (2020) was updated to include the state-of-the-art preprocessing techniques (e.g., BOLD data denoising, single step regression) and newer software versions (Halcomb et al., 2024). The preprocessing was implemented with in-house bash scripts using python 3.6 and FMRIB Software library (FSL version 6.0.1). T1-weighted MPRAGE image of each participant was denoised prior to brain masking and extraction with ANTs (Avants et al., 2009) and then nonlinearly transformed (FSL's flirt and fnirt) to Montreal Neurological Institute (MNI) MNI152 standard space. This MNI-to-T1 transformation was followed by T1-to-EPI transformation, which included linear, nonlinear, and boundary-based registration (Greve & Fischl, 2009) in FSL. This procedure improves registration between the structural masks and functional data and generates standard-to-native (MNI-to-EPI) and inverse (EPI-to-MNI) space transformations required to apply standard space atlases in native EPI space. Functional connectivity data were processed in native EPI space of each participant. The preprocessing included BOLD volume distortion correction using FSL's topup/applytopup (utilizing phase-reversed spin echo field mapping scans), head motion realignment (mcflirt), earlier described T1-to-EPI registration, normalization to mode 1000, and spatial smoothing with a 6 mm isotropic full width at half maximum (FWHM) Gaussian kernel.

Following recommendations for robust preprocessing (Eklund et al., 2016), the preprocessed data were entered into FSL’s MELODIC (Nickerson et al., 2017) for independent components analysis (ICA)-based denoising with ICA-AROMA (Pruim, Mennes, Buitelaar, et al., 2015; Pruim, Mennes, van Rooij, et al., 2015). A single step regression was applied to the denoised BOLD volumes to avoid reintroducing artifacts in the preprocessed denoised data (Lindquist et al., 2019; Phạm et al., 2023). Specifically, regressors were applied that indexed head motion (from the realignment and their derivatives (Power et al., 2014)), accounted for physiological noise (first five signals obtained by PCA from the white matter and cerebrospinal fluid-eroded masks; an implementation of aCompCor (Muschelli et al., 2014)) and three components of the global signal (mean, derivative, squared). Additional regressors performed high-pass filtering (fmin = 0.009 Hz) using Discrete Cosine Transforms bases (Shirer et al., 2015), and included outlier volume despiking (Phạm et al., 2023). The outlier volumes were scrubbed based on the significant “DVARS” metrics obtained on the single-regression preprocessed data as described in (Phạm et al., 2023). The scrubbed volume values were replaced using the linear interpolation of their nearest remaining timepoints. The mean number of scrubbed volumes was 0.5% (Standard Deviation: 0.35%; range: 0 – 1.5%) across both scans for all participants. The number of scrubbed volumes between the two FHA groups did not differ ($p = 0.20$ for the first resting-only fMRI scan, $p = 0.79$ for the second, task-rest scan; two-tailed t-test).

A whole-brain data-driven functional parcellation based on 300 regions, as obtained by (Schaefer et al., 2018) was projected into native EPI space of each participant as described earlier. For the subcortex, we used a 32 node Scale II parcellation (Melbourne subcortical atlas) as defined by Tian et al., (2020), which resulted in 332 total brain regions.

## 2.4 Riemannian Geometry in Functional Connectivity

Functional connectivity is most often estimated pairwise using the Pearson correlation coefficient of two brain regions' BOLD time series, resulting in a symmetric correlation matrix for whole-brain functional connectivity (Fornito et al., 2016). Such symmetric matrices are positive definite if they are invertible (all their eigenvalues are greater than zero). The collection of such matrices forms a non-linear topological space, or manifold, that is referred to as symmetric positive definite (SPD) (You & Park, 2021). If correlation matrices are singular (at least one eigenvalue equal to 0), they are positive semi-definite and lie on the edge of the SPD manifold. This occurs, for instance, when computing FCs if the parcellation has more brain regions than the number of time-points in the BOLD time-series. Such matrices, however, become positive definite through regularization, where their main diagonal entries are incremented by a regularization value (Abbas et al., 2021; Venkatesh et al., 2020).

The canonical methods that use Euclidean or correlation distance of vectorized FCs ignore their topological properties and the interrelatedness between their elements (Abbas et al., 2023; You & Park, 2021). Consequently, comparisons of FCs belonging to the SPD manifold must account for their non-Euclidean geometry by using the Affine Invariant Riemannian Metric (AIRM) (Pennec et al., 2006), also known as geodesic distance. Alternatively, we can apply tangent space projection, where the FCs are projected into a Euclidean space that is tangent to a reference matrix, $C_{ref}$, on the SPD manifold (Pervaiz et al., 2020). For any reference matrix belonging to the SPD manifold, its tangent space is a collection of vectors that are the derivatives of the curves crossing that matrix on the manifold (You & Park, 2021). The geodesic distance between FCs on the manifold can be approximated by the Euclidean distance of their corresponding projections on the tangent space (Barachant et al., 2013). The projection is computed using the formula in Equation 1.

$$\hat{S} = \log_m(C_{ref}^{-\frac{1}{2}} \cdot S \cdot C_{ref}^{-\frac{1}{2}}) \qquad (1)$$

where $\hat{S}$ is the projected matrix on tangent space, $S$ is the matrix on the SPD manifold, $C_{ref}$ is the reference matrix on the manifold, and $\log_m$ is the matrix logarithm function.

Estimated FCs are always positive semidefinite. However, they are not guaranteed to be positive definite and hence may not be invertible. In that case, regularization is an essential step when using geodesic distance or tangent space projection. The choice of regularization value can affect the resulting tangent-FCs, thus impacting fingerprinting accuracy (Abbas et al., 2023) and predictions based on tangent-FCs. The reference matrix choice, $C_{ref}$, can also affect the outcome. The reference matrix, which represents a centroid of the data, is often obtained as a function of all or a subset of FCs in the dataset (Pervaiz et al., 2020).

Abbas et al., (2023) evaluated manifold and tangent-FCs fingerprinting accuracy (Amico & Goñi, 2018; Finn et al., 2015)– an important criterion for the reliability of FCs and the predictions or models based on them. They found that **(1)** the combination of correlation distance and tangent-FCs had the highest fingerprinting accuracy for all Human Connectome Project fMRI conditions and for all parcellation granularities evaluated; **(2)** the optimal regularization value for fingerprinting (Abbas et al., 2021) was consistently 0.01 (the smallest value tested) for the combination of correlation distance and tangent-FCs; **(3)** the Riemann function, defined in Equation 2 (Fletcher et al., 2004; Moakher, 2005), was the best choice to compute the reference, $C_{ref}$, for fingerprinting accuracy.

$$C_{ref} = \arg\min_{C} \sum_{i} d_G\left(\widehat{C_{ref}}, S_i\right)^2 \qquad (2)$$

where $C_{ref}$ is the Riemann mean reference matrix, $S_i$ is the $i$th SPD matrix, $\widehat{C_{ref}}$ is the initial matrix for the computation of $C_{ref}$ (e.g., arithmetic mean of $S_i, \forall i$ (Pervaiz et al., 2020)) and $d_G$ is the geodesic distance function. After projection, the elements of the projected matrices become unrelated features (Ng et al., 2016). We can thus use correlation distance to compare tangent-FCs.

### 2.5 Functional Reconfiguration Analysis

We estimated pairwise functional connectivity of brain regions by calculating the Pearson correlation coefficient of their corresponding preprocessed BOLD time-series. Both scans were divided into 4-minute segments (the SST duration), resulting in one SST segment and four resting state segments (R) and labeled as R1.1, R1.2, SST, R2.1, and R2.2 (Figure 1). Regularized FCs were projected to tangent space using the Riemann mean of R1.1 FCs as the reference using PyRiemann library (Barachant, 2015). We defined functional reconfiguration from changing task demands (from rest-to-task or from task-to-rest) as the distance between their corresponding functional connectomes (SPD FCs or tangent-FCs). Specifically, we measured functional reconfiguration of each participant from R1.2 to SST (engaging the task) and SST to R2.1 (disengaging from the task) by calculating the correlation distance of their corresponding tangent-FCs (Figure 1E).

The engaging and disengaging functional reconfiguration of participants served as the response variable in a multilinear regression model, with sex, age, education, SSRT, CES-D score, recent drinking behavior, and FHA status (0=negative, 1=positive) as predictors. Here, recent drinking behavior was derived as the first principal component of the AUDIT score, and self-reported drinking days, drinks per week, and drinks per drinking day (derived from the timeline follow-back technique (Sobell et al., 1986)), which explained 65% of variance. We performed two separate leave-one-out cross validations to evaluate the stability of the final models (regression coefficients) with respect to (1) the sample variation in the final models and (2) the sample variation in computing the reference, $C_{ref}$, for tangent space projection. We also tested a range of regularization values (0.001 to 10) to assess their effects on the estimation of functional reconfiguration and ultimately the multilinear regression models.

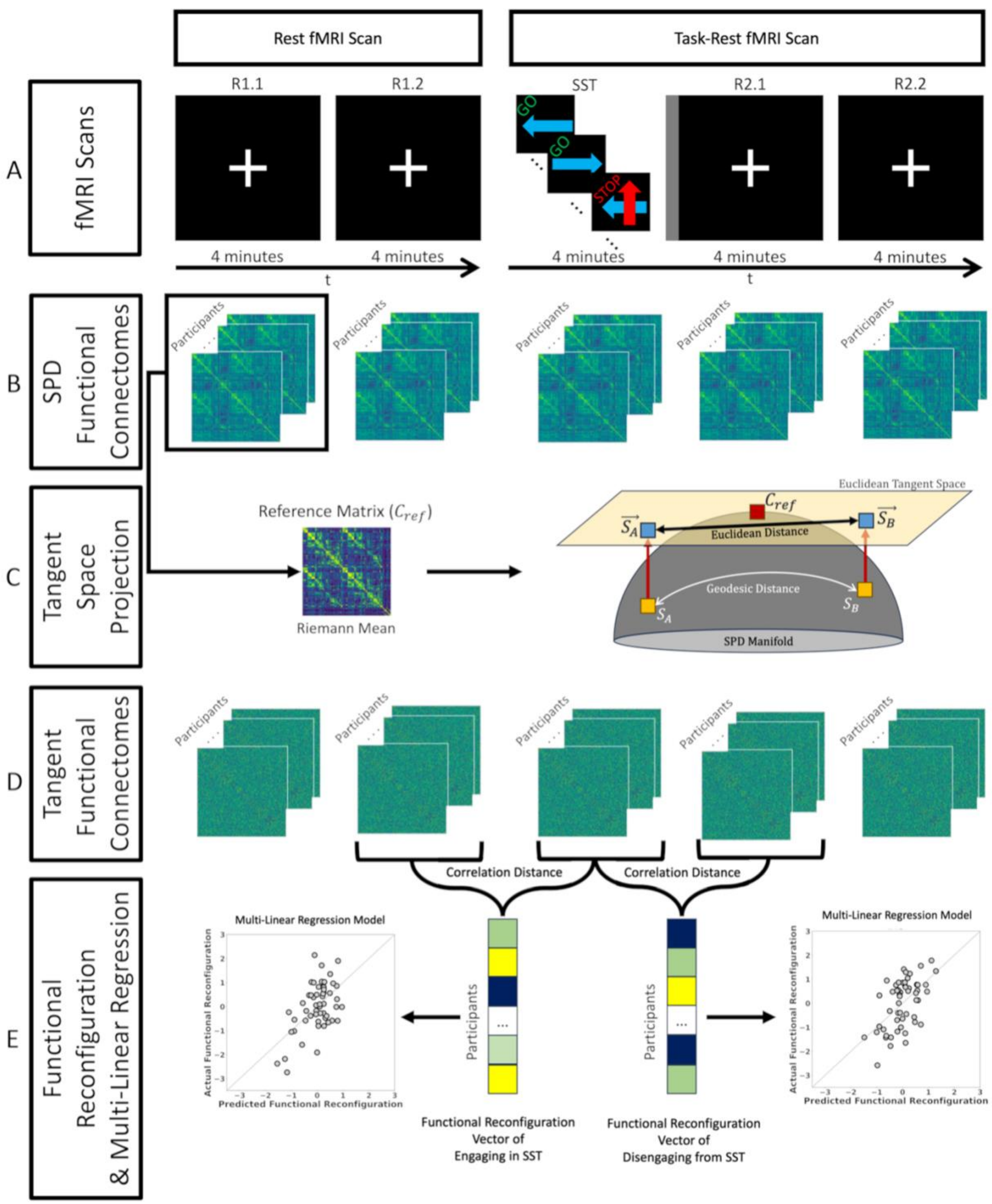

Figure 1. Study design and functional reconfiguration analysis scheme. **(A)** Each participant completed two fMRI scans. Scan 1 (left): 8 minutes of rest with eyes fixated on a central white crosshair; Scan 2 (right): 4-minute stop signal task (GO and STOP labels on top of each trial tile are only illustrative and did not appear on the actual stimuli). The stop signal task segment of the scan was followed by a short 12 s intermission (*gray vertical stripe rectangle*) when a slide announced the upcoming rest with the printed statement, "The task is over. Fix your gaze on the crosshair

for the remainder of the scan". Participants then rested for 8 min and again fixated on the crosshair. **(B)** We divided Scans 1 and 2 into five 4-minute segments (R1.1, R1.2, SST, R2.1, and R2.2) and estimated functional connectivity between brain region pairs using Pearson's correlation coefficient. We subsequently computed Riemann mean of functional connectomes (FCs) from R1.1 **(C)** We used the reference computed in the previous step to project the FCs to tangent space, resulting in their tangent-FCs. **(D)** Functional reconfiguration of each participant from R1.2 to SST and SST to R2.1 were measured as the correlation distance of their corresponding tangent-FCs. We used these functional reconfiguration vectors in two multilinear regression models (engaging and disengaging) where the predictors included family history of alcohol use disorder (FHA), recent drinking behavior, and depressive symptoms (CES-D).

# 3 Results

## 3.1 The choice of fMRI segment for computing $C_{ref}$

After computing the correlation distance of tangent-FCs across all participants for a pair of fMRI segments (e.g., R1.1 and R1.2), we can represent the result as a matrix of size $54 \times 54$, i.e., an *identifiability matrix* (Amico & Goñi, 2018). The diagonal elements of this matrix represent functional reconfiguration of each participant. We computed the identifiability matrices of all pairwise comparisons across fMRI segments (25 matrices) and stacked them to produce a *meta-identifiability matrix* (Figure 2).

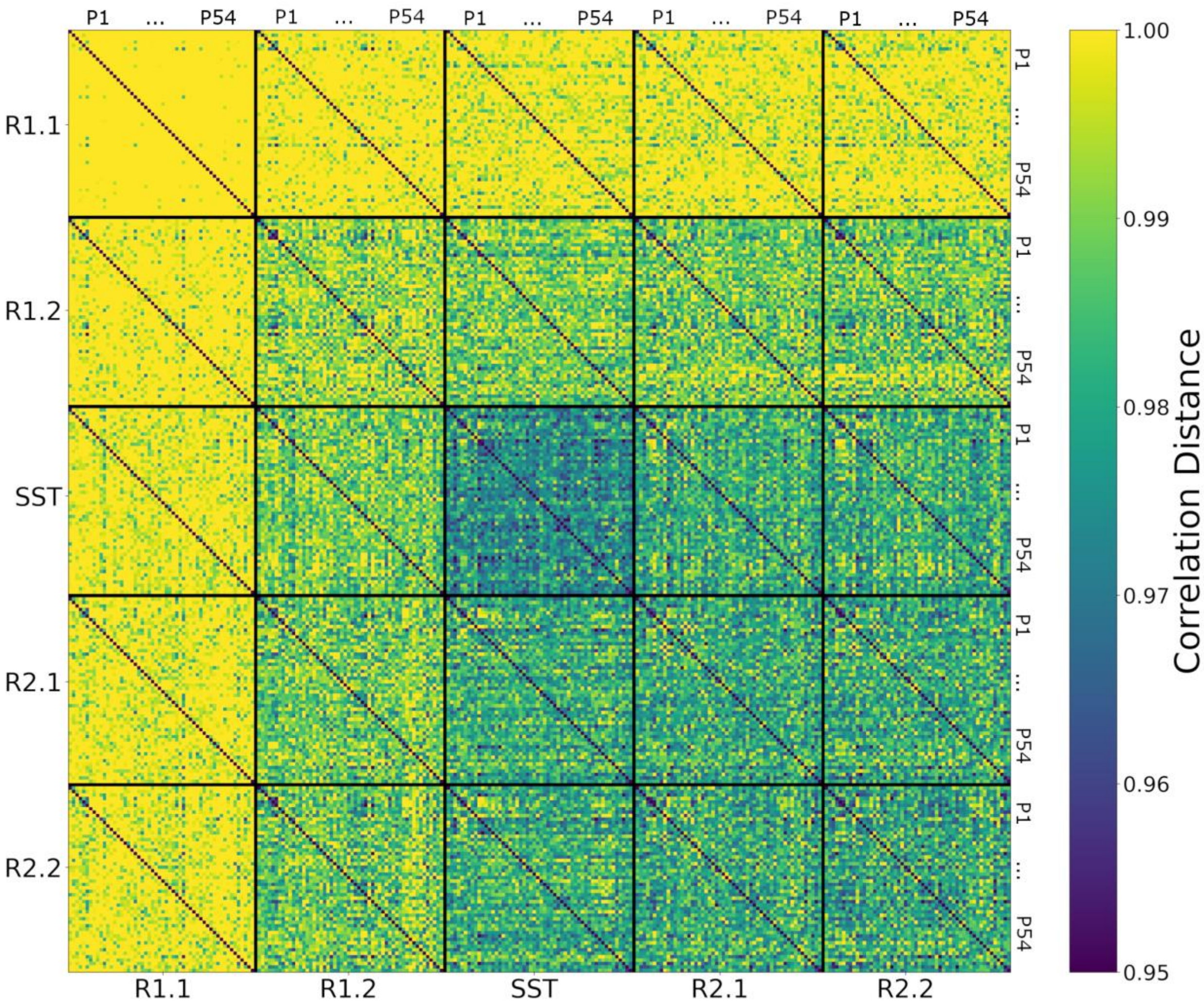

Figure 2. Meta-Identifiability matrix for five fMRI segments consists of 25 identifiability matrices. Each of the five within-segment identifiability matrices is symmetric, contains correlation distances of tangent-FCs across all 54 participants (denoted as P1 … P54), and has zeros on its main diagonal (zero distance between a tangent-FC and itself). Note that identifiability matrices involving different segments are not symmetric. Furthermore, they not only show between-participants distances but also contain, on their main diagonal, functional reconfiguration of each participant from one fMRI segment to another.

As shown in Figure 2, the off-diagonal values of identifiability matrices that involve R1.1 are noticeably larger due to the use of R1.1 manifold FCs in the computation of the reference $C_{ref}$ for tangent space projection. This is because the commonalities of the FCs that are used to compute the reference are cancelled out after projection. Hence, the large correlation distance between them.

This has been described as data *whitening* with respect to the reference in the Riemannian geometry literature (Dadi et al., 2019; Varoquaux et al., 2010).

The same phenomenon appears when using the Riemann mean of R2.2 FCs or the Riemann mean of both R1.1 and R2.2 FCs as the reference matrix, $C_{ref}$, for tangent space projection (Figure 3B and 3C respectively, compared to Figure 3A). Therefore, we evaluated how the use of different resting state fMRI segments to compute the reference, $C_{ref}$, impacts the estimation of within-participant functional reconfiguration (**diagonal elements** of identifiability matrices) and between-participants distances (**off-diagonal elements** of identifiability matrices). Specifically, we separately computed the correlation of diagonal elements (Figures 3D-F) and off-diagonal elements (Figures 3G-I) of each identifiability matrix across pairs of references. Overall, these results indicate that the choice of non-transitioning resting state fMRI segment to compute $C_{ref}$ minimally affects the estimation of the functional reconfiguration values. As shown in Figure 3D-E, the correlation of functional reconfiguration estimates (diagonal elements of each identifiability matrix) using the Riemann mean of R1.1 as the reference and the two alternatives always exceeds 0.96. Note that to preserve the intrinsic characteristics of SST in the tangent-FCs, we only considered only resting state fMRI segments when computing the reference.

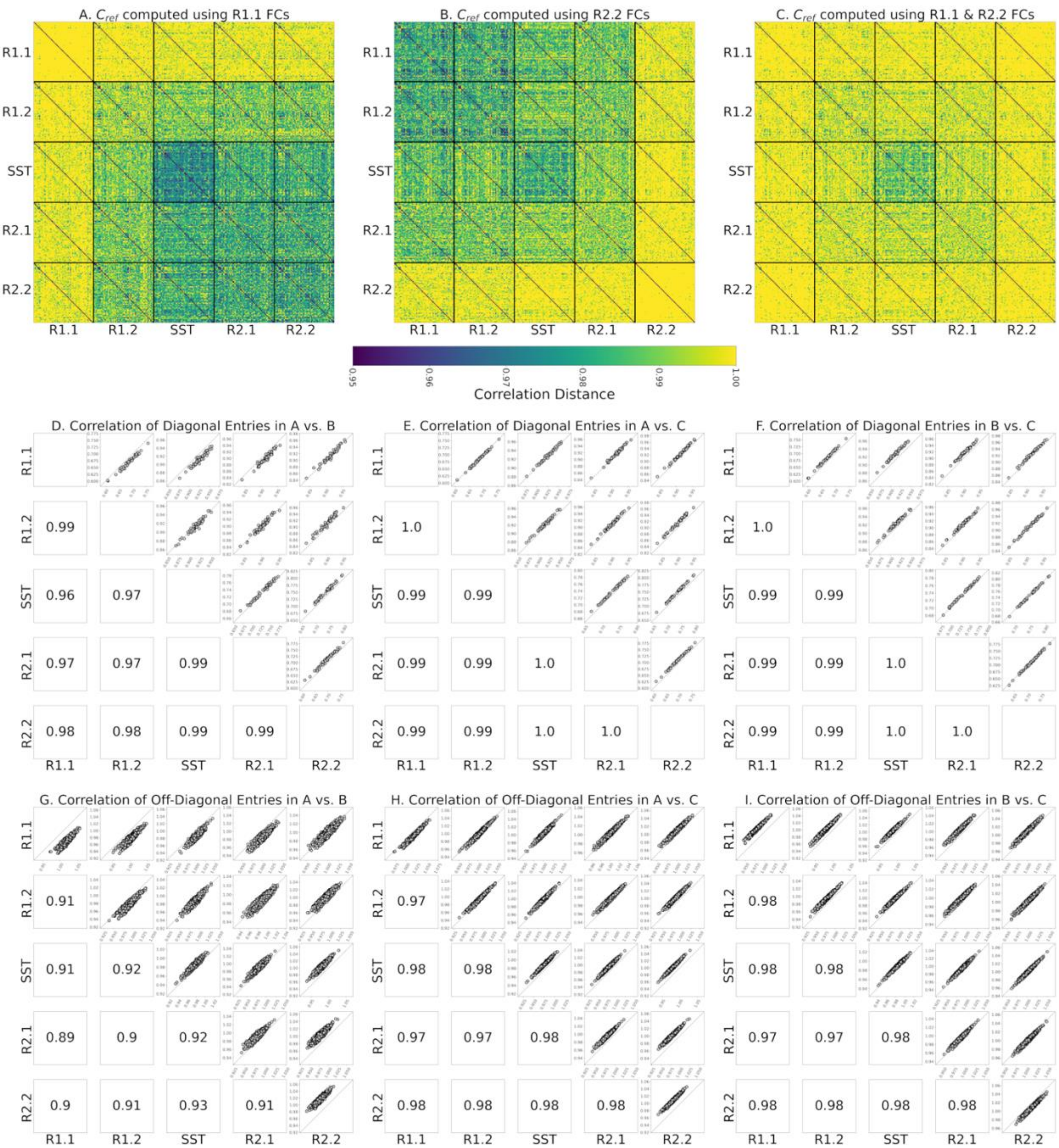

Figure 3. **(A)** Meta-identifiability matrix of tangent-FCs with $C_{ref}$ computed using R1.1 FCs **(B)** Meta-identifiability matrix of tangent-FCs with $C_{ref}$ computed using R2.2 FCs. **(C)** Meta-identifiability matrix of tangent-FCs with $C_{ref}$ computed using R1.1 and R2.2 FCs. The color bar for A, B, and C is the same as Figure 2. **(D)** Correlation of diagonal entries of identifiability matrices that belong to meta-identifiability matrices in A and B. **(E)** Correlation of diagonal entries of identifiability matrices that belong to meta-identifiability matrices in A and C. **(F)** Correlation of diagonal entries of identifiability matrices that belong to meta-identifiability matrices in B and C. **(G)** Correlation of off-diagonal entries of identifiability matrices that belong to meta-identifiability matrices in A and B. The correlations on the main diagonal are 0.90, 0.91, 0.95, 0.90, and 0.92. **(H)** Correlation of diagonal entries of identifiability matrices that belong to meta-identifiability matrices in A and C. Correlations on the main diagonal are 0.97, 0.98, 0.99, 0.97, and 0.98. **(I)** Correlation of diagonal entries of identifiability matrices that belong to meta-identifiability matrices in B and C. Correlations on the main diagonal are 0.98, 0.98, 0.99, 0.97, and 0.98.

## 3.2 Identification rates

In a recent study (Abbas et al., 2023), tangent space projection of FCs improved fingerprinting (measured by identification rate) for all evaluated fMRI conditions and parcellation granularities. The identification rates of this dataset within the blocks of the meta-identifiability matrix are presented before and after tangent space projection, i.e., on the SPD manifold FCs (Figure 4A) and for tangent-FCs (Figure 4B). Near perfect identification rates across two scans only occur with tangent-FCs. This suggests that tangent space projection is mitigating biases due to scan effects.

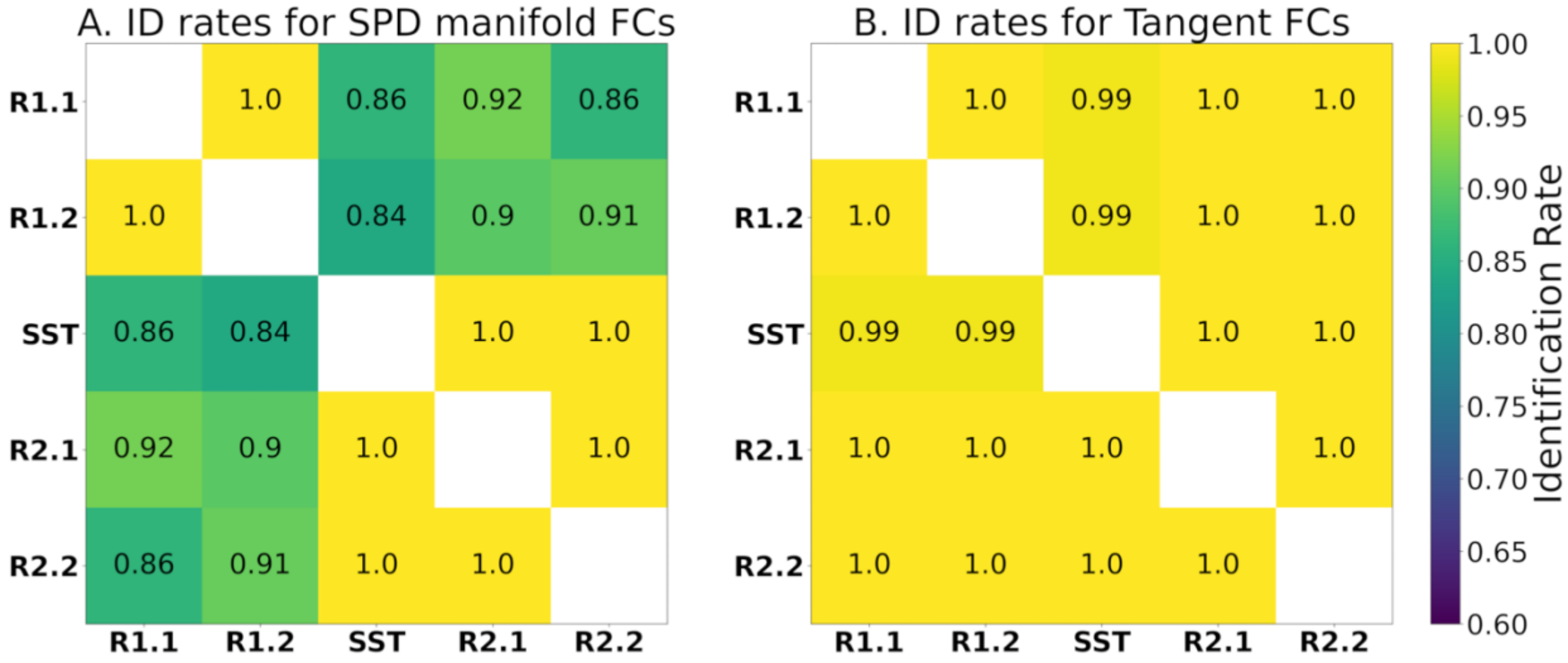

Figure 4. Identification rates of participants across fMRI segments with **(A)** SPD manifold FCs and **(B)** tangent-FCs (regularization value of 0.001 and reference computed using R1.1 FCs). Note that tangent space projection removes potential scan effects in terms of fingerprinting.

## 3.3 Functional Reconfiguration

Functional reconfiguration from rest-to-task (R1.2 to SST), task-to-rest (SST to R2.1) and rest to rest (R2.1 to R2.2, as a baseline control comparison within-rest) are highlighted in Figure 5A. The histograms of functional reconfiguration vectors in Figures 5B.1 to 5B.3 show that individuals functionally reconfigure more when engaging in the SST compared to disengaging from it. Furthermore, the correlation between functional reconfiguration vectors of engaging in and disengaging from the SST is 0.45 (Figure 5C), suggesting that engaging and disengaging from SST for each participant are only partially related. We, therefore, analyzed engaging and disengaging functional reconfiguration with separate multilinear regression models to assess the AUD risk factors associated with each transition (see section 3.4).

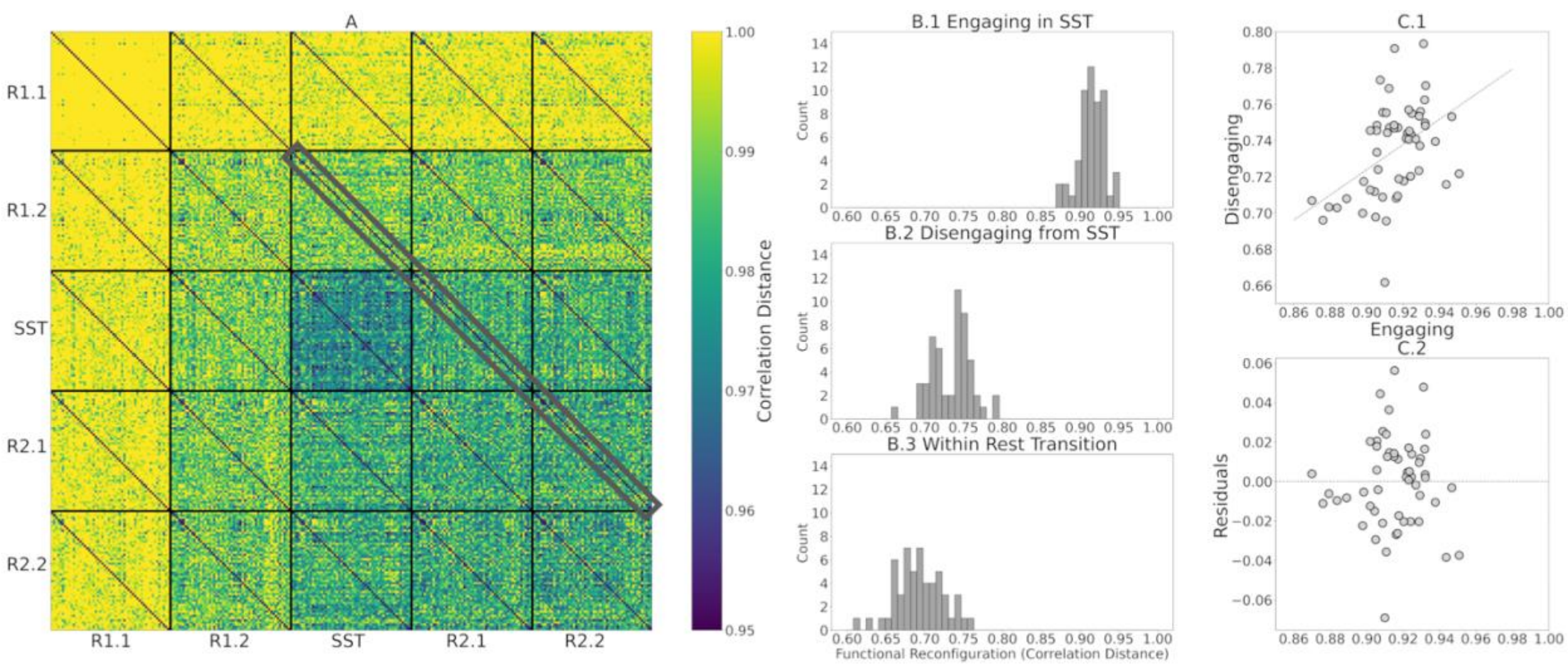

Figure 5. **(A)** Meta-identifiability matrix with functional reconfiguration vectors of R1.2 to SST (engaging), SST to R2.1 (disengaging), and R2.1 to R2.2 (within-rest) highlighted in gray. **(B.1 to B.3)** Histograms of functional reconfiguration values of engaging in the SST, disengaging from the SST, and within-rest post-SST, respectively. We tested functional reconfiguration using paired *T*-tests among Engaging, Disengaging and Within-Rest model and observed significant differences ($p < 0.001$). **(C.1)** Scatter plot of engaging versus disengaging functional reconfiguration vectors, showing that the two are unrelated ($r$=0.44). **(C.2)** scatter plot of residuals when predicting disengaging from engaging.

## 3.4 Associations between functional reconfiguration and AUD risk

We performed three multilinear regression analyses (engaging, disengaging, and within rest) to determine how functional reconfiguration is related to AUD risk. We accounted for possible confounding nuisance variables (age; sex coded as male 1, female 0; education; SSRT on which the FHA groups differed) and related AUD risk variables on which the FHA groups also differed (recent drinking, negative affect/depressive symptoms, assessed by CES-D scores). We first assessed, iteratively, how the cumulative explained variance of the model evolved as the predictors were added in the order described above. This approach showed the contribution of each predictor to the existing model. Subsequently, we built the final models with all seven predictors. Note that the order of the predictors does not affect the results of the final models. The response variable in each model is the functional reconfiguration vector of (1) *engaging in the SST*, i.e., transition from rest-to-task (R1.2 to SST), (2) *disengaging from the SST*, i.e., transition from task-to-rest (SST to R2.1) and (3) *within-rest* transition (R2.1 to R2.2, following the SST) as an intended comparator in which there should be no systematic engagement or disengagement.

When examining functional reconfiguration of *engaging in the* SST, the model explained 31% of the variance (Figure 6A, 1-4). Significant predictors were sex ($p = 0.01$), CES-D ($p = 0.012$) and recent drinking ($p = 0.005$). Being male and higher recent drinking correspond with smaller engaging functional reconfiguration, whereas a higher CES-D score corresponds with greater engaging functional reconfiguration.

Similarly, when examining functional reconfiguration of *disengaging from the SST,* the model explained 31% of the variance (Figure 6B, 1-4). Significant predictors were education ($p = 0.002$), CES-D ($p = 0.028$) and FHA ($p = 0.005$). Higher education and CES-D score correspond with greater disengaging functional reconfiguration, while being FHA positive corresponds with smaller disengaging functional reconfiguration.

Finally, the model for functional reconfiguration of *within-rest transition* explained 21% of the variance (after SST; Figure 6C) and 18% of the variance (before SST; Figure S2A). Note that these two models have less explained variance than engaging and disengaging models. Furthermore, within-rest transitions have significantly smaller magnitudes of functional reconfiguration (Figure 5). The only significant predictor in both models was education ($p = 0.009$ and $p = 0.016$ respectively), with higher education corresponding with greater within-rest functional reconfiguration before and after SST. None of the AUD-risk related variables were significant.

Residual diagnostics for all three models indicate the linear regression assumptions were not violated; see the residual plots in Figures 6A4, 6B4 and 6C4. Finally, when adding head motion (indexed by number of scrubbed volumes across the two scans), it did not change the significance of the other predictors in the multi-linear models, explained little variance by itself, and was not a significant predictor in any of the final models (Figure S3).

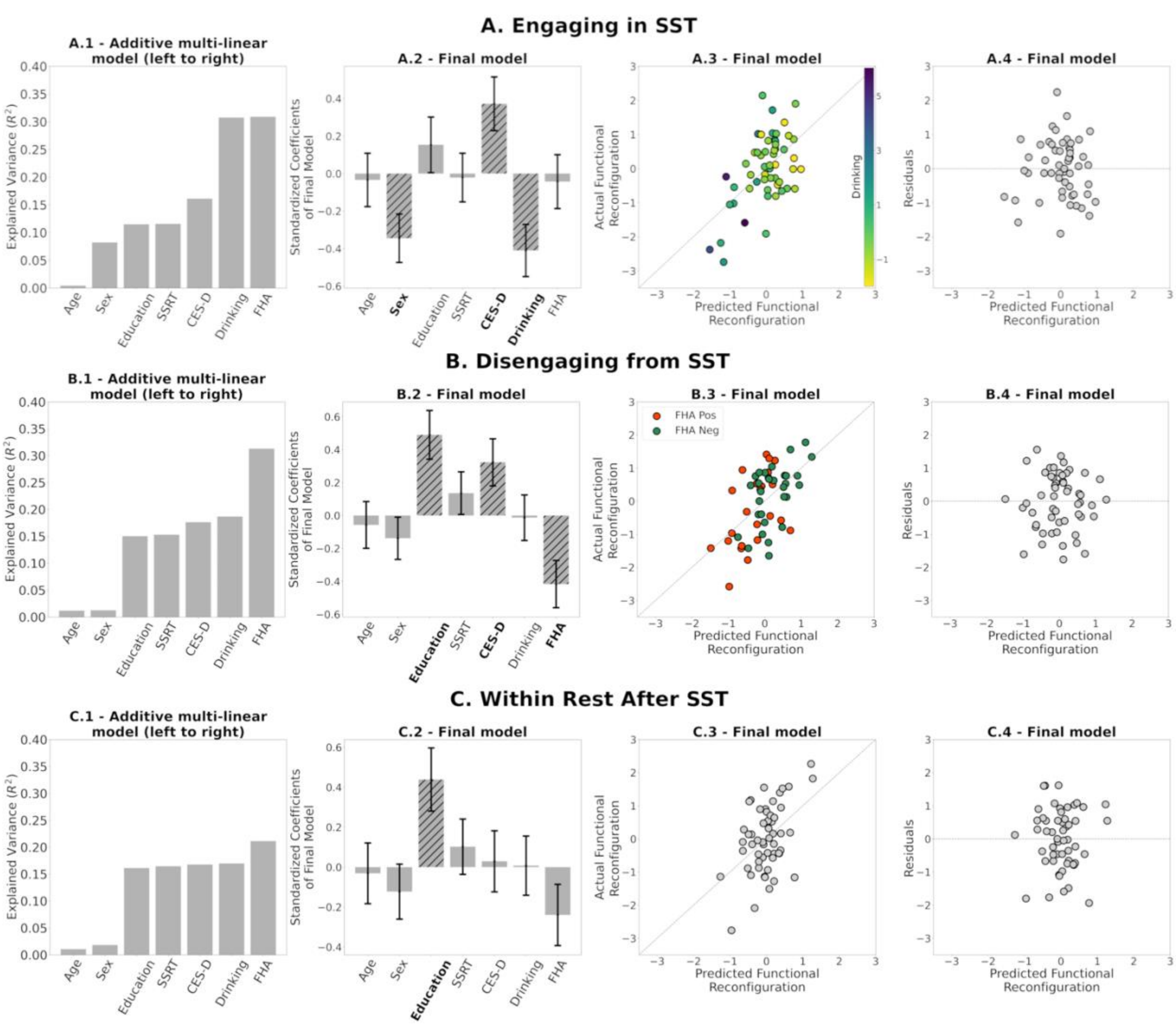

Figure 6. **(A.1, B.1, C.1)** Additive multilinear regression of AUD-risk variables (and adjustment covariates) on functional reconfiguration of engaging in the task (R1.2 to SST), disengaging from the task (SST to R2.1), and during rest after SST (R2.1 to R2.2), with predictors sequentially introduced in the order depicted. See methods for variable definitions **(A.2, B.2, C.2)** Coefficients of the final multilinear regression model, including standard error. Hatched bars and bold labels denote significant predictors ($p \leq 0.05$). **(A.3, B.3, C.3)** Scatter plots of predicted versus actual functional reconfiguration values for the final multilinear regression model in A1, B1, and C1 respectively. Colors in A.3 and B.3 are based on (standardized) recent drinking score and FHA respectively. **(A.4, B.4, C.4)** Scatter plots of the predicted functional reconfiguration versus the standardized residuals of each participant for the final models in **A.1, B.1, C.1** respectively.

### 3.5 Assessing the stability of regression coefficients of final models

To evaluate final models' stability, we performed leave-one-out cross validations for: (1) the multilinear models after using the entire set of SPD FCs of R1.1 to build the reference and project the FCs to tangent space (Figure 7); (2) the computation of the reference for tangent space projection**.** In the latter we project the SPD FCs to tangent space 54 times, each with $C_{ref}$ built on 53 FCs, leaving one participant out. On each iteration, we built a multilinear regression model for engaging in and disengaging from the SST on the projected FCs, which included all 54 participants (Figure 8). The rationale was to understand how much variance in our results comes from the data used for the multilinear models versus the data used to build the reference. The results are presented in Figures 7 and 8, respectively.

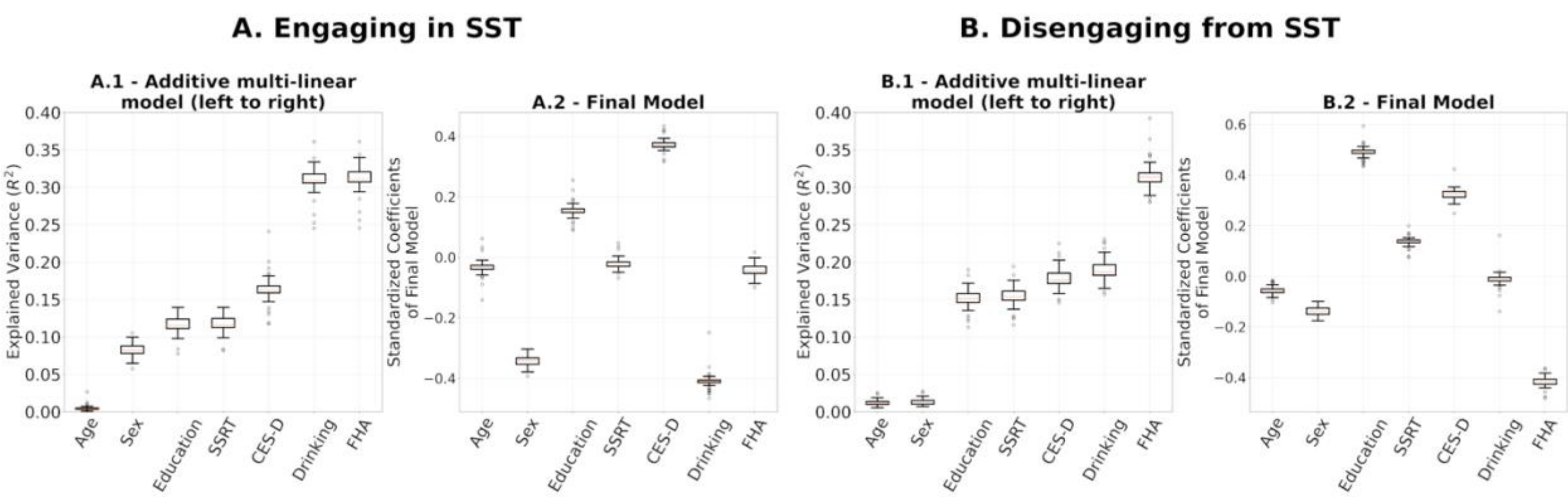

Figure 7. **(A.1, B.1)** Leave-one-out additive multilinear regression model explaining engaging and disengaging functional reconfiguration respectively. **(A.2, B.2)** Distributions of leave-one-out coefficients.

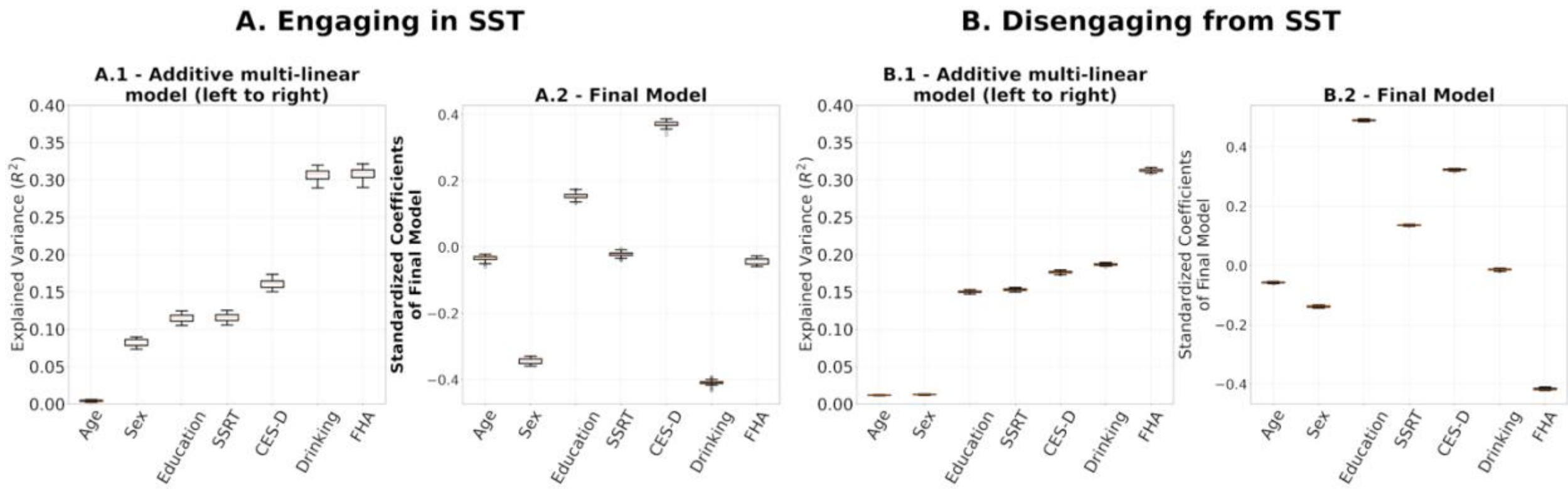

Figure 8. **(A.1, B.1)** Full additive multilinear regression models explaining engaging and disengaging functional reconfiguration respectively with the reference for tangent space projection, $C_{ref}$, built based on leave-one-out. **(A.2, B.2)** Distributions of leave-one-out coefficients.

### 3.6 Regularization Stability

Regularization is essential to ensure positive definiteness of SPD FCs for tangent space projection (see Section 2.4. for details). It is noteworthy that the magnitude of regularization value has an inverse relationship with the range of tangent-FC elements. Moreover, two tangent-FCs derived from the same FC but based on two different regularization values are not guaranteed to be highly correlated (Abbas et al., 2023). This can consequently influence the estimation of functional reconfiguration. Therefore, we tested the performance of the multilinear models presented in Section 3.4 across a range of regularization values (0.0001 to 10). Figure 9 shows that the multilinear regression models explain more variance of functional reconfiguration for regularization values $\leq 0.01$. Furthermore, regularization values also affect the significance of the predictors as shown in Table 3.

When assessing the relationship between regularization value and variance of tangent-FC elements in our dataset, we discovered that for regularization values beyond 0.1, the range and variance of tangent-FC elements become very small, possibly limiting the explanation capacity of our models (Figure S4).

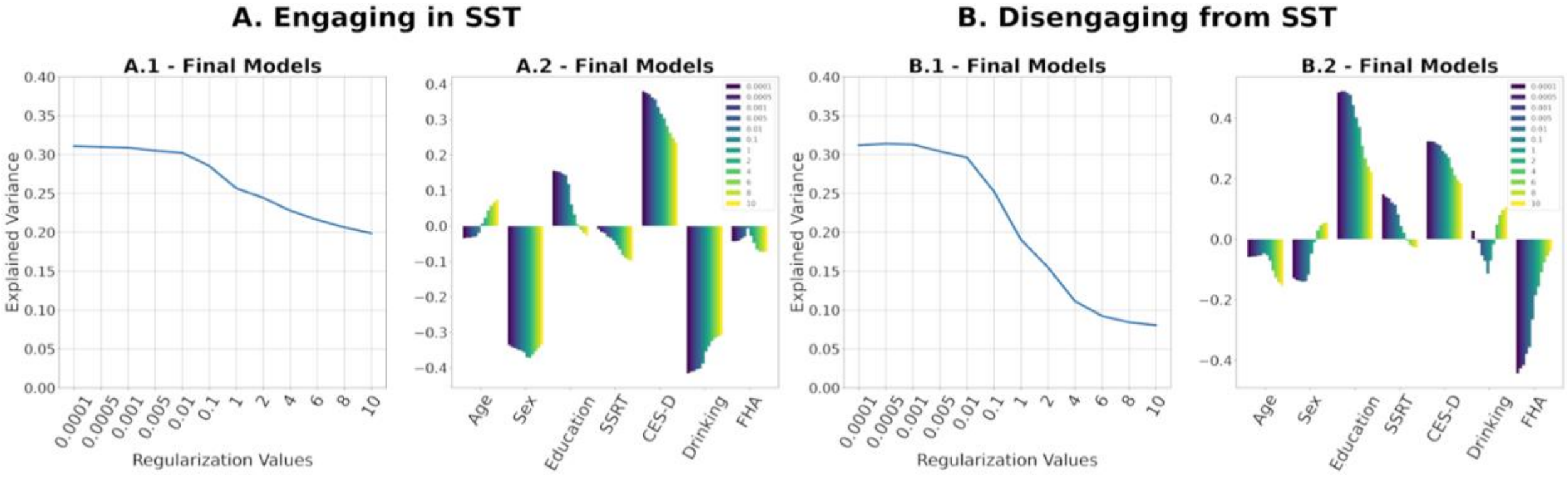

Figure 9. **(A.1, B.1)** Explained variance of multilinear regression models built on tangent-FCs that were projected to tangent space after the regularization of SPD FCs with the regularization values on the x-axis. **(A.2, B.2)** Coefficients of the multilinear regression models, each ordered by the regularization value of their corresponding tangent-FCs.

Table 3. Effect of regularization on the significant predictors of the final engaging and disengaging models (regularization value = 0.001; shown in bold); for every regularization

value, tangent FCs were computed, and functional reconfiguration models were obtained using all seven predictors (see Figure 9 for the effect on regression coefficients)

| | A. Engaging in the SST | | | B. Disengaging from the SST | | |
|---|---|---|---|---|---|---|
| | p-values | | | | | |
| Regularization values | Sex | CES-D | Drinking | Education | CES-D | FHA |
| 0.0001* | 0.012 | 0.010 | 0.004 | 0.002 | 0.027 | 0.003 |
| 0.0005* | 0.010 | 0.011 | 0.005 | 0.002 | 0.027 | 0.004 |
| **0.001*** | **0.010** | **0.012** | **0.005** | **0.002** | **0.028** | **0.005** |
| 0.005* | 0.009 | 0.015 | 0.006 | 0.002 | 0.032 | 0.011 |
| 0.01* | 0.009 | 0.016 | 0.006 | 0.002 | 0.036 | 0.017 |
| 0.1 | 0.009 | 0.025 | 0.008 | 0.006 | 0.055 | 0.082 |
| 1 | 0.007 | 0.038 | 0.017 | 0.015 | 0.073 | 0.237 |
| 2 | 0.008 | 0.048 | 0.023 | 0.028 | 0.094 | 0.327 |
| 4 | 0.010 | 0.069 | 0.031 | 0.071 | 0.151 | 0.502 |
| 6 | 0.013 | 0.090 | 0.037 | 0.121 | 0.203 | 0.641 |
| 8 | 0.016 | 0.112 | 0.041 | 0.164 | 0.241 | 0.744 |
| 10 | 0.020 | 0.133 | 0.044 | 0.196 | 0.267 | 0.815 |

* Regularization values for which all $p \leq 0.05$

# 4 Discussion

In this work, we evaluated the relationship between AUD risk factors and functional reconfiguration both in and from the stop signal task (SST) of motor inhibition. We applied a new methodology to quantify rest-to-task (R1.2 to SST) and task-to-rest (SST to R2.1) functional reconfigurations by calculating the correlation distance of the corresponding tangent-FCs for each participant. Tangent-FCs are the result of applying a Riemannian geometry approach, i.e., tangent space projection (Pennec et al., 2006) to FCs, which projects them to a Euclidean space and transforms their elements into independent features (Abbas et al., 2023; Pervaiz et al., 2020; Tian & Zalesky, 2021; You & Park, 2021). This approach overcomes the limitations of using SPD manifold FCs, specifically interrelatedness of the elements. We also introduced the concept of a meta-identifiability matrix (Figure 2). Given that we divided the scans into five fMRI segments, this symmetric matrix is composed of 25 identifiability matrices. The matrices on the main diagonal correspond with the within fMRI segment comparisons whereas the remaining matrices are comparisons across pairs of fMRI segment. The meta-identifiability matrix depicts the relationship between fMRI segments and participants.

As illustrated in Figure 3A-C, the choice of fMRI segment for computing the reference for tangent space projection affects the magnitude of off-diagonal entries (correlation distance of tangent-FCs across participants) in the meta-identifiability matrix. Specifically, by applying tangent space projection we remove the commonalities of the FCs that are used in the computation of the reference matrix. However, our analysis shows that functional reconfiguration vectors (diagonal elements of each identifiability matrix) across the three meta-identifiability matrices in Figure 3A-C are highly correlated ($\geq 0.96$) (Figure 3D-F); and despite the difference in magnitude of off-diagonal elements, they are also highly correlated across the three computations of the reference matrix ($\geq 0.89$) (Figure 3G-I). Functional reconfiguration estimates are thus stable with respect to the fMRI segments used to compute the reference (Riemann mean of all FCs in that segment).

Recent studies show that tangent space projection improves the predictive power of FCs (Dadi et al., 2019; Dodero et al., 2015; Ng et al., 2017; Qiu et al., 2015; Wong et al., 2018) and fingerprinting accuracy (Abbas et al., 2023), which is also evident in our results (Figure 4). Tangent space projection improved identification rates between segments that belong to different fMRI scans, which suggests that scan effects were minimized. This result comports with recent studies showing how tangent FCs can help harmonize multi-site data (Simeon et al., 2022). This is key regarding our measurements of functional reconfiguration, particularly engaging in the SST, as the transition between rest and task occurred across two scans.

As illustrated by Figure 5B, individuals exhibited more functional reconfiguration when engaging in the SST compared to disengaging from it, both of which are significantly greater than functional reconfiguration of within-rest transition (R2.1 to R2.2) ($p < 0.001$). The difference in the magnitude of functional reconfiguration between engaging in and disengaging from the SST is consistent with previous findings about lingering task effects in subsequent rest periods (Barnes et al., 2009; Chen et al., 2018).

As AUD is highly prevalent (Grant et al., 2015), understanding brain-related vulnerabilities is important to prevention, treatment, and more generally public health, especially given AUD's comorbidity and joint risk with other mental illness (Walters et al., 2018). Prior research of how FHA affects brain connectivity is not extensive, with past work using *a priori* seed regions (Cservenka et al., 2014) or seed-based analyses of data collected during cognitive tasks (Herting et al., 2011; Weiland et al., 2013; Wetherill et al., 2012). These data suggest that FHA may affect functional connectivity of the reward and frontal circuits, as evident from related work (Cservenka, 2016). Broader analyses of whole brain regional network connectivity across the brain from resting state studies are less common, but also suggest altered frontal and dorsal premotor and sensorimotor connectivity between those with and without FHA (Holla et al., 2017; Vaidya et al., 2019).

Functional reconfiguration can be measured at multiple spatial and temporal scales within a variety of frameworks, such as similarity measures of FCs (Schultz & Cole, 2016), whole-brain network structure (Shine & Poldrack, 2018), short term fluctuations of whole-brain FCs (Gonzalez-Castillo et al., 2015), trapping efficiency and exit entropy of functional networks (Duong-Tran et al., 2021), modularity and community detection (Bassett et al., 2011; Betzel et al., 2017). Conceptually, functional reconfiguration can be interpreted as relating to: (1) *efficiency,* wherein less functional reconfiguration indicates more efficient transitions between task and rest (Duong-Tran et al., 2021; Schultz & Cole, 2016). (2) *flexibility*, wherein functional reconfiguration is commensurate to the required behavioral adaption to new task demands, and where less functional reconfiguration reflects greater rigidity in transitions between rest and task (Bassett et al., 2011; Betzel et al., 2017). Using the latter interpretation of functional reconfiguration, Amico et al. (2020) uncovered a transient network reconfiguration process during the rest period after task using an approach that extracted independent components of FCs (Amico et al., 2017) which was diminished in individuals with FHA (as well as males).

In this work, we examined the relationship of FHA and functional reconfiguration using the correlation distance of tangent-FCs. The results with this novel framework show that functional reconfiguration from task-to-rest is affected by family history, with those with FHA had less functional reconfiguration. Greater recent drinking behavior (the first principal component of the AUDIT and self-reported drinking) was also associated with reduced functional reconfiguration in the transition from rest-to-task. These findings indicate that AUD risk factors alter functional reconfiguration both while engaging in and disengaging from the SST.

Specifically, in the model for engaging in the SST, those with lower recent drinking functionally reconfigure more after accounting for age, sex, education, and task performance. Sex and depressive symptoms were also significant predictors. Although it is important to control for potential sex-effects in brain function, men are at greater risk of AUD (Flores-Bonilla & Richardson, 2020; Grant et al., 2015), and reconfigured less when engaging in the SST (Figure 6A). Insofar as negative affect/depression figures prominently in AUD, it was unexpected that higher depression scores on the CES-D inventory would relate to greater functional reconfiguration, which might stem from a greater degree of dynamic functional connectivity fluctuations in depression (Lin et al., 2023).

When disengaging from the SST, we observed anticipated FHA effects (i.e., FHA positive individuals functionally reconfigured less), after controlling for education and CES-D (Figure 6B). Similar to the rest-to-task transition, CES-D score was again positively related with the task-to-rest reconfiguration. Note that CES-D was included to control for depression symptoms but, for the entire cohort, mean and standard deviation are 6 and 5 respectively, and where only three subjects are above the clinical threshold (16). The score of FHA positive participants was, however, significantly greater, and even a limited symptom burden might be sufficient to affect the transition between cognitive states.

It has been suggested that better task performance is related to more efficient (smaller) functional reconfiguration (Schultz & Cole, 2016). From that perspective, it could be expected that SSRT (performance measure of the Stop Signal Task) is associated with functional reconfiguration during engaging (and possibly during disengaging). However, SSRT was not significantly associated with functional reconfiguration when engaging or disengaging in our final models (Figure 6). This could be consequence of a limited sample size and/or because SSRT is associated with functional reconfiguration of specific functional circuits rather than of the whole-brain.

The results further show that transitions between the task and rest are asymmetric in that the reconfiguration is more prominent during engagement as compared to disengagement (Figure 5B). We evaluated this observation with respect to the choice of fMRI segments to compute the reference matrix. Our results suggest that the magnitudes of functional reconfiguration were consistent for both engaging and disengaging across the three choices of fMRI segments for computing the reference (Figure S1).

Furthermore, when we measure functional reconfiguration from SST to R2.2 instead of R2.1 as the predicted variable, the multilinear regression model results were similar to those displayed in Figure 6B (i.e., the same predictors were significant, there were only minor differences in the explained variance; Figure S2B). Thus, the effects of FHA group membership on our quantification of functional reconfiguration were not transient but instead spanned the entire rest period after SST.

In contrast to task-rest and rest-task transitions, AUD risk variables were not significant in within-rest transitions after the SST, consistent with our hypotheses. The only significant variable in the final model was education. This suggests that the within-rest fluctuations of functional connectivity following task engagement were unaffected by the stop signal task. Interestingly, the model for within-rest transition in the first scan (R1.1 to R1.2) was also similar to Figure 6C with education again as the only significant predictor (Figure S2A).

We performed leave-one-out cross-validation analyses to study the stability of the regression coefficient estimation in the multilinear regression models and the computation of reference matrix, $C_{ref}$, for tangent space projection. The results in Figures 7 and 8 show that the models presented in Figure 6 were not driven by a particular choice of $C_{ref}$ and were stable across the 54 leave-one-out models fitted. While these results show high stability on the regression coefficients of the final models within our cohort, a larger sample is required to assess the generalizability of our results.

Finally, we evaluated the effects of regularization parameter on the multi-linear regression models. While identification rates were high (>98%) for all tested values, Figure 9 clearly shows that when using correlation distance on tangent-FCs, smaller regularization values best preserve the explanation power of tangent-FCs compared to larger values ($\geq 0.1$). This is due to the *shrinking effect* of regularization on the values of functional couplings in tangent-FCs; the higher the regularization value, the smaller the range and variance of functional couplings (Abbas et al., 2021) (Figure S4). The loss of variance leads to the loss of any meaningful differences between functional couplings. Therefore, depending on the dataset, we recommend testing a range of regularization values starting with the smallest that ensure positive definiteness. Note that although amount of regularization does not affect fingerprinting accuracy of tangent-FCs when using correlation distance (Abbas et al., 2023), our results show that it greatly influences accuracy when explaining functional reconfiguration (Figure 9).

In this dataset reflective of AUD risk, we showed that functional reconfiguration from rest to SST (engaging) and SST to rest (disengaging) measured using tangent-FCs was associated with AUD risk factors. Recent drinking behavior is associated with engaging functional reconfiguration; those with greater recent drinking behavior showed less functional reconfiguration engaging in SST. On the other hand, disengaging from the SST was affected by FHA group membership; FHA positive individuals had less functional reconfiguration while disengaging from the SST. Furthermore, our findings suggest that small values (provided positive definiteness depending on the dataset) preserve the variance of functional couplings, thus preserving the explanation power of tangent-FCs. We aim to expand on the functional reconfiguration framework proposed here by including fMRI designs with rest period before task within the same scan. Our results suggest that analysis of functional reconfiguration using tangent-FCs is a promising avenue to better understand rest-task and task-rest brain transitions.

## 5 Acknowledgments

This work was supported NIH CTSI CTR EPAR2169, NIH R21 AA029614, NIH R01 AA029607, and Indiana Alcohol Research Center P60AA07611, NINDS R01NS126449, NINDS R01NS112303. This research was also supported in part by Lilly Endowment, Inc., through its support for the Indiana University Pervasive Technology Institute and by Shared University Research grants from IBM, Inc., to Indiana University. We would like to thank Drs. Evgeny Chumin and Andrea Avena-Koenigsberger for their invaluable help in testing of the functional analysis pipeline and image visualization. We greatly appreciate Dr. Yu-Chien Wu in The Indiana Institute of Biomedical Imaging Sciences (IIBIS) In-Vivo Imaging Core, for assistance with sequence development and testing, as well as MRI technologists Michele Dragoo, Traci Day, and Robert Bryant for their help during imaging sessions. Finally, we greatly appreciate work by all technicians in Dr. David Kareken's laboratory.

# 6 References

Abbas, K., Liu, M., Venkatesh, M., Amico, E., Kaplan, A. D., Ventresca, M., Pessoa, L., Harezlak, J., & Goñi, J. (2021). Geodesic Distance on Optimally Regularized Functional Connectomes Uncovers Individual Fingerprints. *Brain Connectivity*, *11*(5). https://doi.org/10.1089/brain.2020.0881

Abbas, K., Liu, M., Wang, M., Duong-Tran, D., Tipnis, U., Amico, E., Kaplan, A. D., Dzemidzic, M., Kareken, D., Ances, B. M., Harezlak, J., & Goñi, J. (2023). Tangent functional connectomes uncover more unique phenotypic traits. *IScience*, *26*(9). https://doi.org/10.1016/j.isci.2023.107624

Amico, E., Dzemidzic, M., Oberlin, B. G., Carron, C. R., Harezlak, J., Goñi, J., & Kareken, D. A. (2020). The disengaging brain: Dynamic transitions from cognitive engagement and alcoholism risk. *NeuroImage*, *209*. https://doi.org/10.1016/j.neuroimage.2020.116515

Amico, E., & Goñi, J. (2018). The quest for identifiability in human functional connectomes. *Scientific Reports*, *8*(1). https://doi.org/10.1038/s41598-018-25089-1

Amico, E., Marinazzo, D., Di Perri, C., Heine, L., Annen, J., Martial, C., Dzemidzic, M., Kirsch, M., Bonhomme, V., Laureys, S., & Goñi, J. (2017). Mapping the functional connectome traits of levels of consciousness. *NeuroImage*, *148*. https://doi.org/10.1016/j.neuroimage.2017.01.020

Avants, B., Tustison, N., & Johnson, H. (2009). Advanced Normalization Tools (ANTS). *Insight Journal*.

Band, G. P. H., van der Molen, M. W., & Logan, G. D. (2003). Horse-race model simulations of the stop-signal procedure. *Acta Psychologica*, *112*(2). https://doi.org/10.1016/S0001-6918(02)00079-3

Barachant, A. (2015). PyRiemann: Python package for covariance matrices manipulation and Biosignal classification with application in Brain Computer interface, 2015. *URL Https://Github. Com/Alexandrebarachant/PyRiemann*.

Barachant, A., Bonnet, S., Congedo, M., & Jutten, C. (2013). Classification of covariance matrices using a Riemannian-based kernel for BCI applications. *Neurocomputing*, *112*. https://doi.org/10.1016/j.neucom.2012.12.039

Bari, S., Amico, E., Vike, N., Talavage, T. M., & Goñi, J. (2019). Uncovering multi-site identifiability based on resting-state functional connectomes. *NeuroImage*, *202*. https://doi.org/10.1016/j.neuroimage.2019.06.045

Barnes, A., Bullmore, E. T., & Suckling, J. (2009). Endogenous human brain dynamics recover slowly following cognitive effort. *PLoS ONE*, *4*(8). https://doi.org/10.1371/journal.pone.0006626

Bassett, D. S., Wymbs, N. F., Porter, M. A., Mucha, P. J., Carlson, J. M., & Grafton, S. T. (2011). Dynamic reconfiguration of human brain networks during learning. *Proceedings of the National Academy of Sciences of the United States of America*, *108*(18). https://doi.org/10.1073/pnas.1018985108

Betzel, R. F., Satterthwaite, T. D., Gold, J. I., & Bassett, D. S. (2017). Positive affect, surprise, and fatigue are correlates of network flexibility. *Scientific Reports*, *7*(1). https://doi.org/10.1038/s41598-017-00425-z

Braun, U., Schäfer, A., Walter, H., Erk, S., Romanczuk-Seiferth, N., Haddad, L., Schweiger, J. I., Grimm, O., Heinz, A., Tost, H., Meyer-Lindenberg, A., & Bassett, D. S. (2015). Dynamic reconfiguration of frontal brain networks during executive cognition in humans.

*Proceedings of the National Academy of Sciences of the United States of America*, *112*(37). https://doi.org/10.1073/pnas.1422487112

Bucholz, K. K., Cadoret, R., Cloninger, C. R., Dinwiddie, S. H., Hesselbrock, V. M., Nurnberger, J. I., Reich, T., Schmidt, I., & Schuckit, M. A. (1994). A new, semi-structured psychiatric interview for use in genetic linkage studies: A report on the reliability of the SSAGA. *Journal of Studies on Alcohol*, *55*(2). https://doi.org/10.15288/jsa.1994.55.149

Buckner, R. L., Krienen, F. M., & Yeo, B. T. T. (2013). Opportunities and limitations of intrinsic functional connectivity MRI. In *Nature Neuroscience* (Vol. 16, Issue 7). https://doi.org/10.1038/nn.3423

Chen, R. H., Ito, T., Kulkarni, K. R., & Cole, M. W. (2018). The Human Brain Traverses a Common Activation-Pattern State Space Across Task and Rest. *Brain Connectivity*, *8*(7). https://doi.org/10.1089/brain.2018.0586

Cservenka, A. (2016). Neurobiological phenotypes associated with a family history of alcoholism. In *Drug and Alcohol Dependence* (Vol. 158). https://doi.org/10.1016/j.drugalcdep.2015.10.021

Cservenka, A., Casimo, K., Fair, D. A., & Nagel, B. J. (2014). Resting state functional connectivity of the nucleus accumbens in youth with a family history of alcoholism. *Psychiatry Research - Neuroimaging*, *221*(3). https://doi.org/10.1016/j.pscychresns.2013.12.004

Dadi, K., Rahim, M., Abraham, A., Chyzhyk, D., Milham, M., Thirion, B., & Varoquaux, G. (2019). Benchmarking functional connectome-based predictive models for resting-state fMRI. *NeuroImage*, *192*. https://doi.org/10.1016/j.neuroimage.2019.02.062

Dodero, L., Minh, H. Q., Biagio, M. S., Murino, V., & Sona, D. (2015). Kernel-based classification for brain connectivity graphs on the Riemannian manifold of positive definite matrices. *Proceedings - International Symposium on Biomedical Imaging*, *2015-July*. https://doi.org/10.1109/ISBI.2015.7163812

Duong-Tran, D., Abbas, K., Amico, E., Corominas-Murtra, B., Dzemidzic, M., Kareken, D., Ventresca, M., & Goñi, J. (2021). A morphospace of functional configuration to assess configural breadth based on brain functional networks. *Network Neuroscience*, *5*(3). https://doi.org/10.1162/netn_a_00193

Eagle, D. M., Baunez, C., Hutcheson, D. M., Lehmann, O., Shah, A. P., & Robbins, T. W. (2008). Stop-signal reaction-time task performance: Role of prefrontal cortex and subthalamic nucleus. *Cerebral Cortex*, *18*(1). https://doi.org/10.1093/cercor/bhm044

Eichenbaum, A., Pappas, I., Cohen, D. L. J. R., & D'esposito, M. (2021). Differential contributions of static and time-varying functional connectivity to human behavior. *Network Neuroscience*, *5*(1). https://doi.org/10.1162/netn_a_00172

Eklund, A., Nichols, T. E., & Knutsson, H. (2016). Cluster failure: Why fMRI inferences for spatial extent have inflated false-positive rates. *Proceedings of the National Academy of Sciences of the United States of America*, *113*(28). https://doi.org/10.1073/pnas.1602413113

Finn, E. S. (2021). Is it time to put rest to rest? In *Trends in Cognitive Sciences* (Vol. 25, Issue 12). https://doi.org/10.1016/j.tics.2021.09.005

Finn, E. S., & Bandettini, P. A. (2021). Movie-watching outperforms rest for functional connectivity-based prediction of behavior. *NeuroImage*, *235*. https://doi.org/10.1016/j.neuroimage.2021.117963

Finn, E. S., Scheinost, D., Finn, D. M., Shen, X., Papademetris, X., & Constable, R. T. (2017). Can brain state be manipulated to emphasize individual differences in functional connectivity? In *NeuroImage* (Vol. 160). https://doi.org/10.1016/j.neuroimage.2017.03.064

Finn, E. S., Shen, X., Scheinost, D., Rosenberg, M. D., Huang, J., Chun, M. M., Papademetris, X., & Constable, R. T. (2015). Functional connectome fingerprinting: Identifying individuals using patterns of brain connectivity. *Nature Neuroscience*, *18*(11). https://doi.org/10.1038/nn.4135

Fletcher, P. T., Lu, C., Pizer, S. M., & Joshi, S. (2004). Principal geodesic analysis for the study of nonlinear statistics of shape. *IEEE Transactions on Medical Imaging*, *23*(8). https://doi.org/10.1109/TMI.2004.831793

Flores-Bonilla, A., & Richardson, H. N. (2020). Sex differences in the neurobiology of alcohol use disorder. *Alcohol Research: Current Reviews*, *40*(2). https://doi.org/10.35946/arcr.v40.2.04

Fornito, A., Zalesky, A., & Breakspear, M. (2015). The connectomics of brain disorders. In *Nature Reviews Neuroscience* (Vol. 16, Issue 3). https://doi.org/10.1038/nrn3901

Fornito, A., Zalesky, A., & Bullmore, E. T. (2016). Fundamentals of Brain Network Analysis. In *Fundamentals of Brain Network Analysis*. https://doi.org/10.1016/B978-0-12-407908-3.09996-9

Gallen, C. L., Turner, G. R., Adnan, A., & D'Esposito, M. (2016). Reconfiguration of brain network architecture to support executive control in aging. *Neurobiology of Aging*, *44*. https://doi.org/10.1016/j.neurobiolaging.2016.04.003

Gonzalez-Castillo, J., Hoy, C. W., Handwerker, D. A., Robinson, M. E., Buchanan, L. C., Saad, Z. S., & Bandettini, P. A. (2015). Tracking ongoing cognition in individuals using brief, whole-brain functional connectivity patterns. *Proceedings of the National Academy of Sciences of the United States of America*, *112*(28). https://doi.org/10.1073/pnas.1501242112

Grant, B. F., Goldstein, R. B., Saha, T. D., Patricia Chou, S., Jung, J., Zhang, H., Pickering, R. P., June Ruan, W., Smith, S. M., Huang, B., & Hasin, D. S. (2015). Epidemiology of DSM-5 alcohol use disorder results from the national epidemiologic survey on alcohol and related conditions III. *JAMA Psychiatry*, *72*(8). https://doi.org/10.1001/jamapsychiatry.2015.0584

Greene, A. S., Gao, S., Scheinost, D., & Constable, R. T. (2018). Task-induced brain state manipulation improves prediction of individual traits. *Nature Communications*, *9*(1). https://doi.org/10.1038/s41467-018-04920-3

Greve, D. N., & Fischl, B. (2009). Accurate and robust brain image alignment using boundary-based registration. *NeuroImage*, *48*(1). https://doi.org/10.1016/j.neuroimage.2009.06.060

Halcomb, M. E., Dzemidzic, M., Avena-Koenigsberger, A., Hile, K. L., Durazzo, T. C., & Yoder, K. K. (2024). Greater ventral striatal functional connectivity in cigarette smokers relative to non-smokers across a spectrum of alcohol consumption. *Brain Imaging and Behavior*. https://doi.org/10.1007/s11682-024-00903-9

Hearne, L. J., Mattingley, J. B., & Cocchi, L. (2016). Functional brain networks related to individual differences in human intelligence at rest. *Scientific Reports*, *6*. https://doi.org/10.1038/srep32328

Herting, M. M., Fair, D., & Nagel, B. J. (2011). Altered fronto-cerebellar connectivity in alcohol-naïve youth with a family history of alcoholism. *NeuroImage*, *54*(4). https://doi.org/10.1016/j.neuroimage.2010.10.030

Holla, B., Panda, R., Venkatasubramanian, G., Biswal, B., Bharath, R. D., & Benegal, V. (2017). Disrupted resting brain graph measures in individuals at high risk for alcoholism.

*Psychiatry Research - Neuroimaging*, *265*. https://doi.org/10.1016/j.pscychresns.2017.05.002

Jiang, R., Zuo, N., Ford, J. M., Qi, S., Zhi, D., Zhuo, C., Xu, Y., Fu, Z., Bustillo, J., Turner, J. A., Calhoun, V. D., & Sui, J. (2020). Task-induced brain connectivity promotes the detection of individual differences in brain-behavior relationships. *NeuroImage*, *207*. https://doi.org/10.1016/j.neuroimage.2019.116370

Li, N., Ma, N., Liu, Y., He, X. S., Sun, D. L., Fu, X. M., Zhang, X., Han, S., & Zhang, D. R. (2013). Resting-state functional connectivity predicts impulsivity in economic decision-making. *Journal of Neuroscience*, *33*(11). https://doi.org/10.1523/JNEUROSCI.1342-12.2013

Lin, X., Jing, R., Chang, S., Liu, L., Wang, Q., Zhuo, C., Shi, J., Fan, Y., Lu, L., & Li, P. (2023). Understanding the heterogeneity of dynamic functional connectivity patterns in first-episode drug naïve depression using normative models. *Journal of Affective Disorders*, *327*. https://doi.org/10.1016/j.jad.2023.01.109

Lindquist, M. A., Geuter, S., Wager, T. D., & Caffo, B. S. (2019). Modular preprocessing pipelines can reintroduce artifacts into fMRI data. *Human Brain Mapping*, *40*(8). https://doi.org/10.1002/hbm.24528

Moakher, M. (2005). A differential geometric approach to the geometric mean of symmetric positive-definite matrices. *SIAM Journal on Matrix Analysis and Applications*, *26*(3). https://doi.org/10.1137/S0895479803436937

Muschelli, J., Nebel, M. B., Caffo, B. S., Barber, A. D., Pekar, J. J., & Mostofsky, S. H. (2014). Reduction of motion-related artifacts in resting state fMRI using aCompCor. *NeuroImage*, *96*. https://doi.org/10.1016/j.neuroimage.2014.03.028

Ng, B., Varoquaux, G., Poline, J. B., Greicius, M., & Thirion, B. (2016). Transport on Riemannian manifold for connectivity-based brain decoding. *IEEE Transactions on Medical Imaging*, *35*(1). https://doi.org/10.1109/TMI.2015.2463723

Ng, B., Varoquaux, G., Poline, J. B., Thirion, B., Greicius, M. D., & Poston, K. L. (2017). Distinct alterations in Parkinson's medication-state and disease-state connectivity. *NeuroImage: Clinical*, *16*. https://doi.org/10.1016/j.nicl.2017.09.004

Nickerson, L. D., Smith, S. M., Öngür, D., & Beckmann, C. F. (2017). Using dual regression to investigate network shape and amplitude in functional connectivity analyses. *Frontiers in Neuroscience*, *11*(MAR). https://doi.org/10.3389/fnins.2017.00115

Nurnberger, J. I., Wiegand, R., Bucholz, K., O'Connor, S., Meyer, E. T., Reich, T., Rice, J., Schuckit, M., King, L., Petti, T., Bierut, L., Hinrichs, A. L., Kuperman, S., Hesselbrock, V., & Porjesz, B. (2004). A family study of alcohol dependence: Coaggregation of multiple disorders in relatives of alcohol-dependent probands. *Archives of General Psychiatry*, *61*(12). https://doi.org/10.1001/archpsyc.61.12.1246

Pennec, X., Fillard, P., & Ayache, N. (2006). A riemannian framework for tensor computing. *International Journal of Computer Vision*, *66*(1). https://doi.org/10.1007/s11263-005-3222-z

Pervaiz, U., Vidaurre, D., Woolrich, M. W., & Smith, S. M. (2020). Optimising network modelling methods for fMRI. *NeuroImage*, *211*. https://doi.org/10.1016/j.neuroimage.2020.116604

Phạm, D., McDonald, D. J., Ding, L., Nebel, M. B., & Mejia, A. F. (2023). Less is more: balancing noise reduction and data retention in fMRI with data-driven scrubbing. *NeuroImage*, *270*. https://doi.org/10.1016/j.neuroimage.2023.119972

Power, J. D., Mitra, A., Laumann, T. O., Snyder, A. Z., Schlaggar, B. L., & Petersen, S. E. (2014). Methods to detect, characterize, and remove motion artifact in resting state fMRI. *NeuroImage*, *84*. https://doi.org/10.1016/j.neuroimage.2013.08.048

Pruim, R. H. R., Mennes, M., Buitelaar, J. K., & Beckmann, C. F. (2015). Evaluation of ICA-AROMA and alternative strategies for motion artifact removal in resting state fMRI. *NeuroImage*, *112*. https://doi.org/10.1016/j.neuroimage.2015.02.063

Pruim, R. H. R., Mennes, M., van Rooij, D., Llera, A., Buitelaar, J. K., & Beckmann, C. F. (2015). ICA-AROMA: A robust ICA-based strategy for removing motion artifacts from fMRI data. *NeuroImage*, *112*. https://doi.org/10.1016/j.neuroimage.2015.02.064

Qiu, A., Lee, A., Tan, M., & Chung, M. K. (2015). Manifold learning on brain functional networks in aging. *Medical Image Analysis*, *20*(1). https://doi.org/10.1016/j.media.2014.10.006

Radloff, L. S. (1977). The CES-D Scale. *Applied Psychological Measurement*, *1*(3). https://doi.org/10.1177/014662167700100306

Rosenberg, M. D., Finn, E. S., Scheinost, D., Papademetris, X., Shen, X., Constable, R. T., & Chun, M. M. (2015). A neuromarker of sustained attention from whole-brain functional connectivity. *Nature Neuroscience*, *19*(1). https://doi.org/10.1038/nn.4179

SAUNDERS, J. B., AASLAND, O. G., BABOR, T. F., DE LA FUENTE, J. R., & GRANT, M. (1993). Development of the Alcohol Use Disorders Identification Test (AUDIT): WHO Collaborative Project on Early Detection of Persons with Harmful Alcohol Consumption-II. *Addiction*, *88*(6). https://doi.org/10.1111/j.1360-0443.1993.tb02093.x

Schaefer, A., Kong, R., Gordon, E. M., Laumann, T. O., Zuo, X.-N., Holmes, A. J., Eickhoff, S. B., & Yeo, B. T. T. (2018). Local-Global Parcellation of the Human Cerebral Cortex from Intrinsic Functional Connectivity MRI. *Cerebral Cortex*, *28*(9). https://doi.org/10.1093/cercor/bhx179

Schultz, D. H., & Cole, M. W. (2016). Higher intelligence is associated with less task-related brain network reconfiguration. *Journal of Neuroscience*, *36*(33). https://doi.org/10.1523/JNEUROSCI.0358-16.2016

Shine, J. M., Breakspear, M., Bell, P. T., Ehgoetz Martens, K., Shine, R., Koyejo, O., Sporns, O., & Poldrack, R. A. (2019). Human cognition involves the dynamic integration of neural activity and neuromodulatory systems. *Nature Neuroscience*, *22*(2). https://doi.org/10.1038/s41593-018-0312-0

Shine, J. M., & Poldrack, R. A. (2018). Principles of dynamic network reconfiguration across diverse brain states. In *NeuroImage* (Vol. 180). https://doi.org/10.1016/j.neuroimage.2017.08.010

Shirer, W. R., Jiang, H., Price, C. M., Ng, B., & Greicius, M. D. (2015). Optimization of rs-fMRI Pre-processing for Enhanced Signal-Noise Separation, Test-Retest Reliability, and Group Discrimination. *NeuroImage*, *117*. https://doi.org/10.1016/j.neuroimage.2015.05.015

Simeon, G., Piella, G., Camara, O., & Pareto, D. (2022). Riemannian Geometry of Functional Connectivity Matrices for Multi-Site Attention-Deficit/Hyperactivity Disorder Data Harmonization. *Frontiers in Neuroinformatics*, *16*. https://doi.org/10.3389/fninf.2022.769274

Smith, S. M., Jenkinson, M., Woolrich, M. W., Beckmann, C. F., Behrens, T. E. J., Johansen-Berg, H., Bannister, P. R., De Luca, M., Drobnjak, I., Flitney, D. E., Niazy, R. K., Saunders, J., Vickers, J., Zhang, Y., De Stefano, N., Brady, J. M., & Matthews, P. M. (2004). Advances in functional and structural MR image analysis and implementation as FSL. *NeuroImage*, *23*(SUPPL. 1). https://doi.org/10.1016/j.neuroimage.2004.07.051

Smitha, K. A., Akhil Raja, K., Arun, K. M., Rajesh, P. G., Thomas, B., Kapilamoorthy, T. R., & Kesavadas, C. (2017). Resting state fMRI: A review on methods in resting state connectivity analysis and resting state networks. In *Neuroradiology Journal* (Vol. 30, Issue 4). https://doi.org/10.1177/1971400917697342

Sobell, M. B., Sobell, L. C., Klajner, F., Pavan, D., & Basian, E. (1986). The reliability of a timeline method for assessing normal drinker college students' recent drinking history: Utility for alcohol research. *Addictive Behaviors*, *11*(2). https://doi.org/10.1016/0306-4603(86)90040-7

Svaldi, D. O., Goñi, J., Abbas, K., Amico, E., Clark, D. G., Muralidharan, C., Dzemidzic, M., West, J. D., Risacher, S. L., Saykin, A. J., & Apostolova, L. G. (2021). Optimizing differential identifiability improves connectome predictive modeling of cognitive deficits from functional connectivity in Alzheimer's disease. *Human Brain Mapping*, *42*(11). https://doi.org/10.1002/hbm.25448

Tian, Y., Margulies, D. S., Breakspear, M., & Zalesky, A. (2020). Topographic organization of the human subcortex unveiled with functional connectivity gradients. *Nature Neuroscience*, *23*(11). https://doi.org/10.1038/s41593-020-00711-6

Tian, Y., & Zalesky, A. (2021). Machine learning prediction of cognition from functional connectivity: Are feature weights reliable? *NeuroImage*, *245*. https://doi.org/10.1016/j.neuroimage.2021.118648

Tung, K. C., Uh, J., Mao, D., Xu, F., Xiao, G., & Lu, H. (2013). Alterations in resting functional connectivity due to recent motor task. *NeuroImage*, *78*. https://doi.org/10.1016/j.neuroimage.2013.04.006

Vaidya, J. G., Elmore, A. L., Wallace, A. L., Langbehn, D. R., Kramer, J. R., Kuperman, S., & O'Leary, D. S. (2019). Association Between Age and Familial Risk for Alcoholism on Functional Connectivity in Adolescence. *Journal of the American Academy of Child and Adolescent Psychiatry*, *58*(7). https://doi.org/10.1016/j.jaac.2018.12.008

Van Der Wijk, G., Harris, J. K., Hassel, S., Davis, A. D., Zamyadi, M., Arnott, S. R., Milev, R., Lam, R. W., Frey, B. N., Hall, G. B., Müller, D. J., Rotzinger, S., Kennedy, S. H., Strother, S. C., Macqueen, G. M., & Protzner, A. B. (2022). Baseline Functional Connectivity in Resting State Networks Associated with Depression and Remission Status after 16 Weeks of Pharmacotherapy: A CAN-BIND Report. *Cerebral Cortex*, *32*(6). https://doi.org/10.1093/cercor/bhab286

Varoquaux, G., Baronnet, F., Kleinschmidt, A., Fillard, P., & Thirion, B. (2010). Detection of brain functional-connectivity difference in post-stroke patients using group-level covariance modeling. *Lecture Notes in Computer Science (Including Subseries Lecture Notes in Artificial Intelligence and Lecture Notes in Bioinformatics)*, *6361 LNCS*(PART 1). https://doi.org/10.1007/978-3-642-15705-9_25

Venkatesh, M., Jaja, J., & Pessoa, L. (2020). Comparing functional connectivity matrices: A geometry-aware approach applied to participant identification. *NeuroImage*, *207*. https://doi.org/10.1016/j.neuroimage.2019.116398

Walters, R. K., Polimanti, R., Johnson, E. C., McClintick, J. N., Adams, M. J., Adkins, A. E., Aliev, F., Bacanu, S. A., Batzler, A., Bertelsen, S., Biernacka, J. M., Bigdeli, T. B., Chen, L. S., Clarke, T. K., Chou, Y. L., Degenhardt, F., Docherty, A. R., Edwards, A. C., Fontanillas, P., … Agrawal, A. (2018). Transancestral GWAS of alcohol dependence reveals common genetic underpinnings with psychiatric disorders. *Nature Neuroscience*, *21*(12). https://doi.org/10.1038/s41593-018-0275-1

Weiland, B. J., Welsh, R. C., Yau, W. Y. W., Zucker, R. A., Zubieta, J. K., & Heitzeg, M. M. (2013). Accumbens functional connectivity during reward mediates sensation-seeking and alcohol use in high-risk youth. *Drug and Alcohol Dependence*, *128*(1–2). https://doi.org/10.1016/j.drugalcdep.2012.08.019

Wetherill, R. R., Bava, S., Thompson, W. K., Boucquey, V., Pulido, C., Yang, T. T., & Tapert, S. F. (2012). Frontoparietal connectivity in substance-naïve youth with and without a family history of alcoholism. *Brain Research*, *1432*. https://doi.org/10.1016/j.brainres.2011.11.013

Wong, E., Anderson, J. S., Zielinski, B. A., & Fletcher, P. T. (2018). Riemannian regression and classification models of brain networks applied to autism. *Lecture Notes in Computer Science (Including Subseries Lecture Notes in Artificial Intelligence and Lecture Notes in Bioinformatics)*, *11083 LNCS*. https://doi.org/10.1007/978-3-030-00755-3_9

You, K., & Park, H. J. (2021). Re-visiting Riemannian geometry of symmetric positive definite matrices for the analysis of functional connectivity. *NeuroImage*, *225*. https://doi.org/10.1016/j.neuroimage.2020.117464

Zhao, W., Makowski, C., Hagler, D. J., Garavan, H. P., Thompson, W. K., Greene, D. J., Jernigan, T. L., & Dale, A. M. (2023). Task fMRI paradigms may capture more behaviorally relevant information than resting-state functional connectivity. *NeuroImage*, *270*. https://doi.org/10.1016/j.neuroimage.2023.119946

## Supplemental Material

Table S1. Loadings of recent drinking behavior principal components

| | PC 1 | PC 2 |
|---|---|---|
| Loadings | | |
| AUDIT | 0.90 | 0.03 |
| Total drinking days | 0.68 | -0.71 |
| Drinks per week | 0.97 | -0.02 |
| Drinks per drinking day | 0.63 | 0.75 |
| Eigenvalues | 2.63 | 1.08 |
| Explained Variance | 64% | 26% |

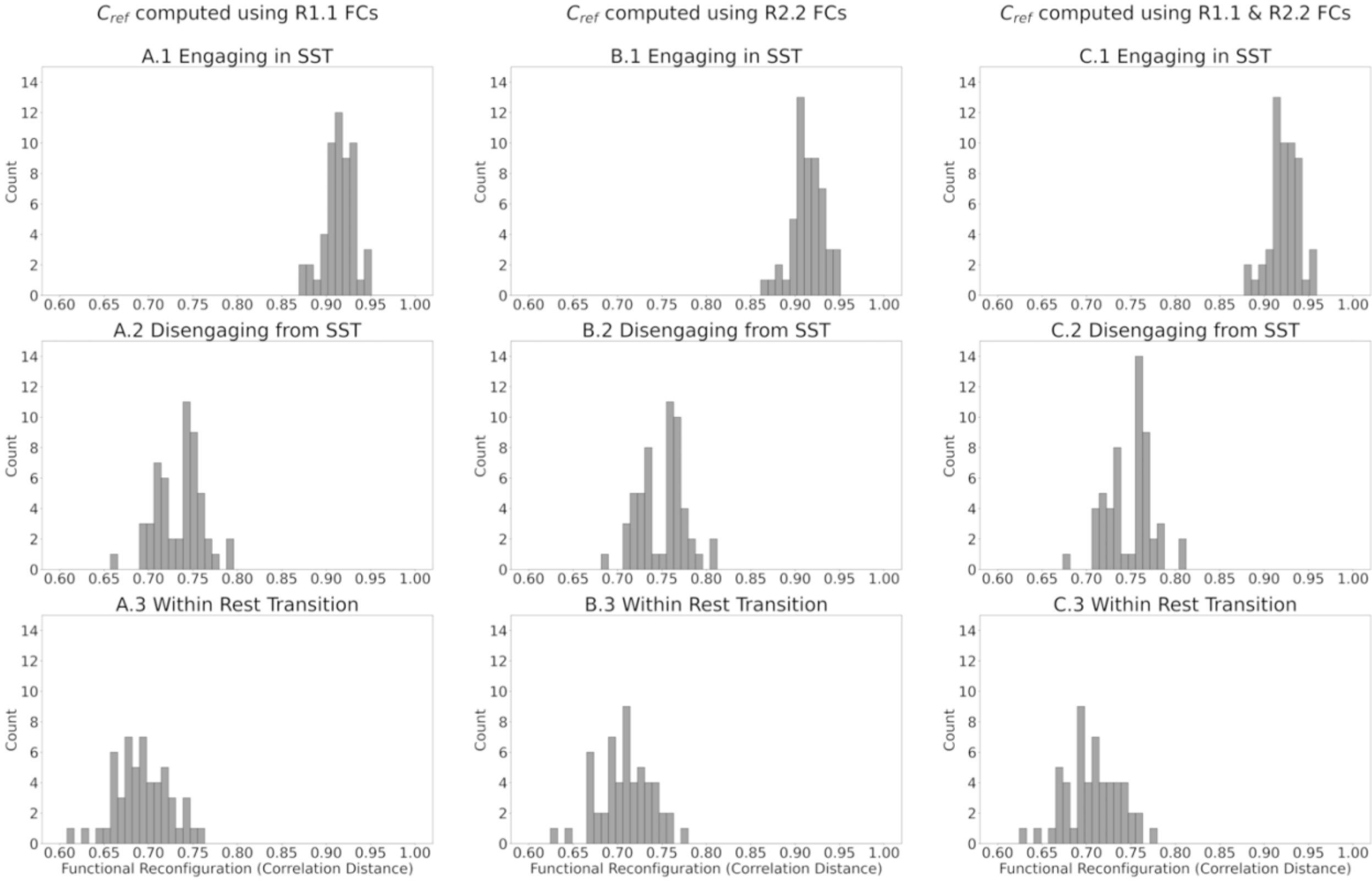

Figure S1. **(A1-A3)** Histograms of functional reconfiguration values of engaging to SST, disengaging from the SST, and within post-SST rest with $C_{ref}$ as the Riemann mean of R1.1 FCs. **(B1-B3)** Histograms of functional reconfiguration values of engaging to SST, disengaging from the SST, and within post-SST rest with $C_{ref}$ as the Riemann mean of R2.2 FCs. **(C1-C3)** Histograms of functional reconfiguration values of engaging to SST, disengaging from the SST, and within post-SST rest with $C_{ref}$ as the Riemann mean of R1.1 and R2.2 FCs.

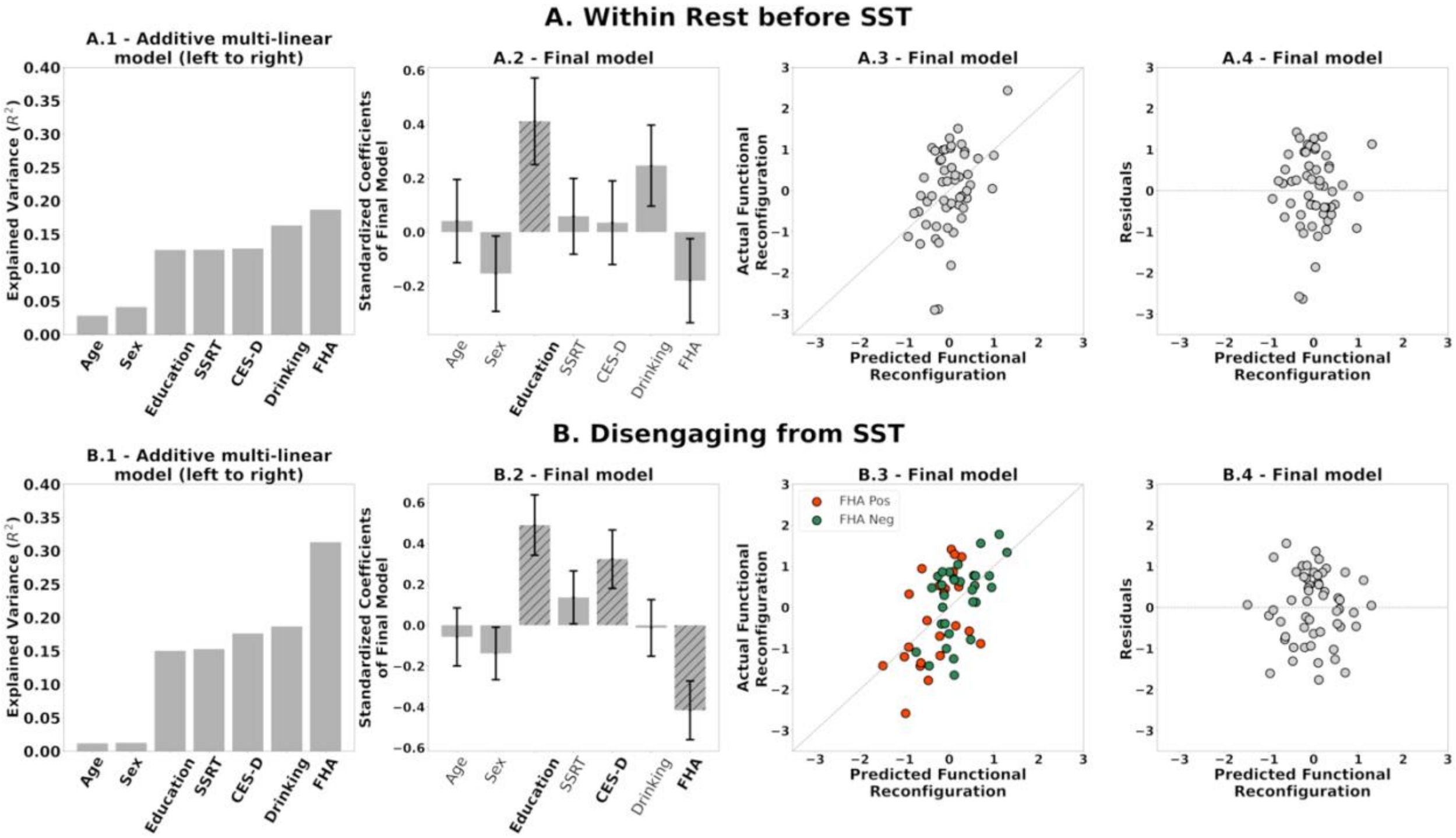

Figure S2. **(A.1, B.1)** Additive multilinear regression of AUD-risk variables (and adjustment covariates) on functional reconfiguration of within-rest before SST (R1.1 to R1.2) and disengaging from SST using the second rest segment after SST (SST to R2.2) with predictors sequentially introduced in the order depicted. See methods for variable definitions **(A.2, B.2)** Coefficients of the final multilinear regression model, including standard error. Hatched bars and bold labels denote significant predictors ($p \leq 0.05$). **(A.3, B.3)** Scatter plots of predicted versus actual functional reconfiguration values for the final multilinear regression model in A1, B1, and C1 respectively. Colors in B.3 are based on FHA status. **(A.4, B.4)** Scatter plots of the predicted functional reconfiguration versus the standardized residuals of each participant for the final models in **A.1, B.1** respectively.

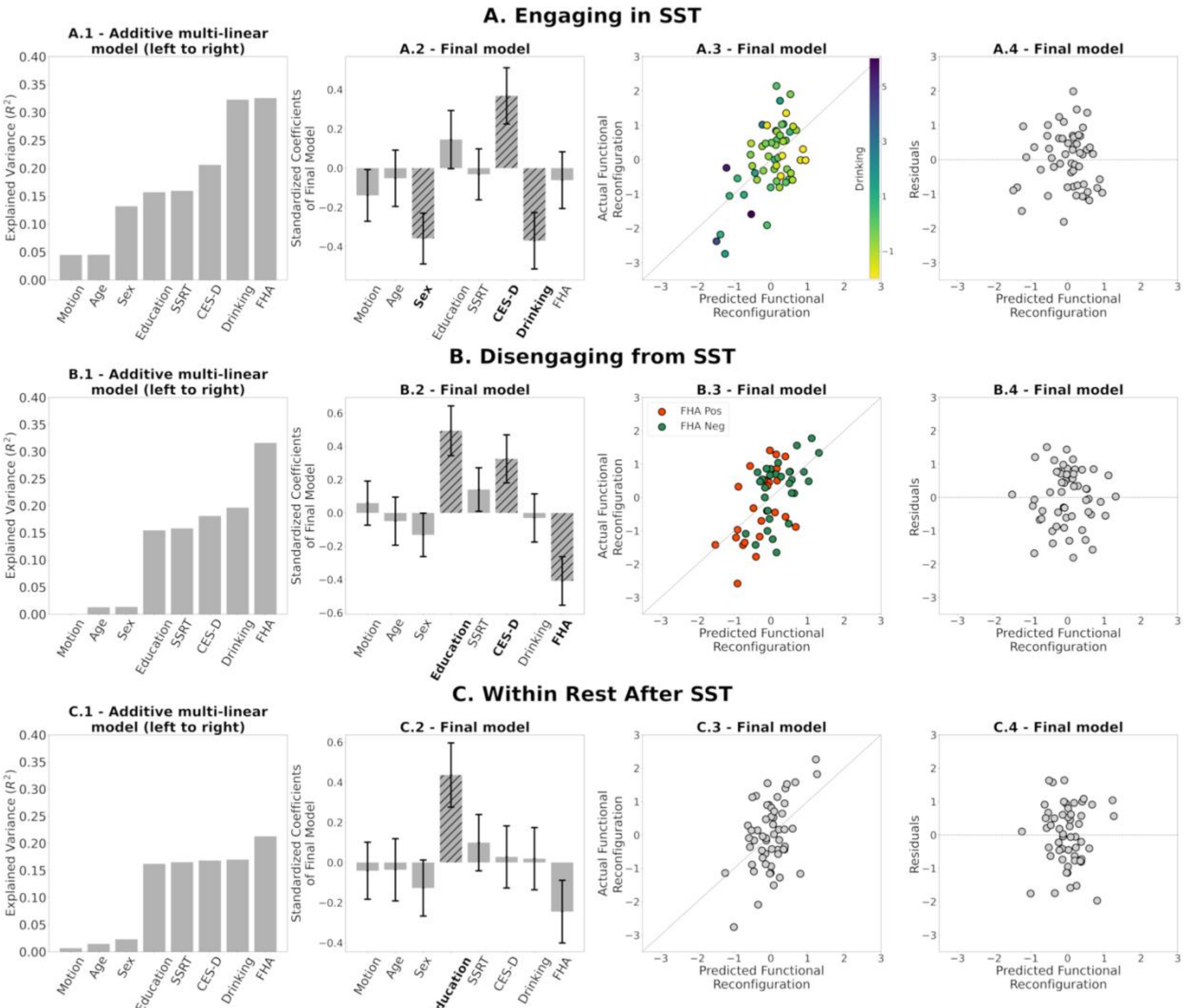

Figure S3. **(A.1, B.1, C.1)** Additive multilinear regression of AUD-risk variables (and adjustment covariates) on functional reconfiguration of engaging in the task (R1.2 to SST), disengaging from the task (SST to R2.1), and during rest after SST (R2.1 to R2.2), with predictors sequentially introduced in the order depicted. See methods for variable definitions **(A.2, B.2, C.2)** Coefficients of the final multilinear regression model, including standard error. Hatched bars and bold labels denote significant predictors ($p \leq 0.05$). **(A.3, B.3, C.3)** Scatter plots of predicted versus actual functional reconfiguration values for the final multilinear regression model in A1, B1, and C1 respectively. Colors in A.3 and B.3 are based on (standardized) recent drinking score and FHA respectively. **(A.4, B.4, C.4)** Scatter plots of the predicted functional reconfiguration versus the standardized residuals of each participant for the final models in **A.1, B.1, C.1** respectively.

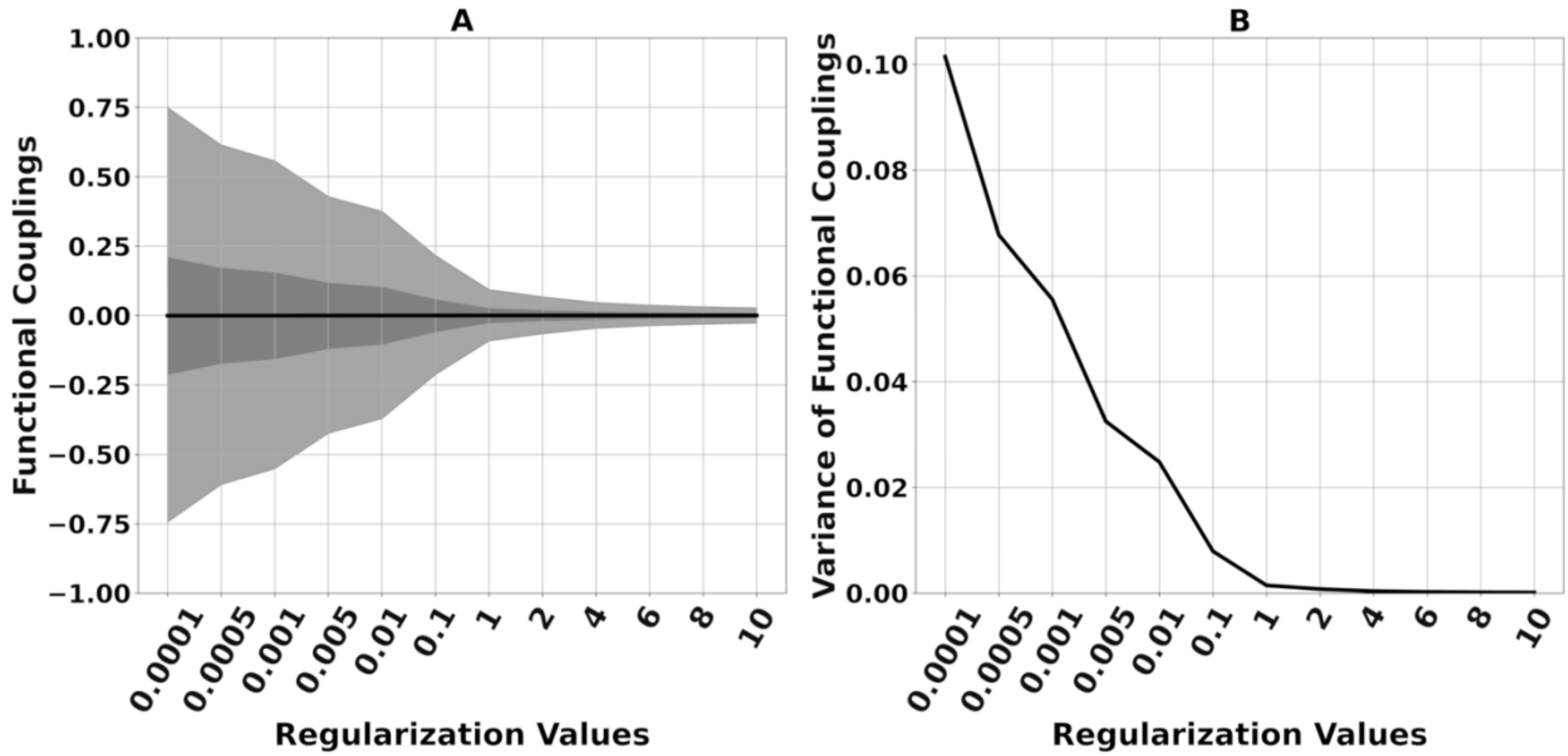

Figure S4. **(A)** Upper and lower bounds of functional couplings (elements of tangent-FCs) for each regularization value indicated by the shaded area spanning the 1st to 99th percentiles of functional couplings of all tangent FCs. The darker region shows the 25th to 75th percentile range. The black line indicates the mean. **(B)** Variance of functional couplings for each regularization value.